\documentclass[english,aps,showpacs,superscriptaddress,floatfix, notitlepage,prl,longbibliography,reprint,unicode=true,colorlinks=true,citecolor=Blue,linkcolor=RubineRed,urlcolor=Blue]{revtex4-1}
\usepackage[T1]{fontenc}
\usepackage[latin9]{inputenc}
\usepackage{babel}
\usepackage{amsmath}
\usepackage{amsthm}
\usepackage{amssymb}
\usepackage{graphicx}
\usepackage[unicode=true]
 {hyperref}

\makeatletter
\theoremstyle{plain}
\newtheorem{thm}{\protect\theoremname}
\theoremstyle{plain}
\newtheorem{prop}{\protect\propositionname}

\usepackage{braket}
\usepackage{bm}
\usepackage{txfonts} 
\usepackage{graphicx}
\usepackage[usenames,dvipsnames]{xcolor}
\date{\today}

\makeatother

\providecommand{\propositionname}{Proposition}
\providecommand{\theoremname}{Theorem}

\begin{document}
\title{Quantum Measurement Encoding for Quantum Metrology}
\author{Jing Yang~\href{https://orcid.org/0000-0002-3588-0832}{\includegraphics[scale=0.05]{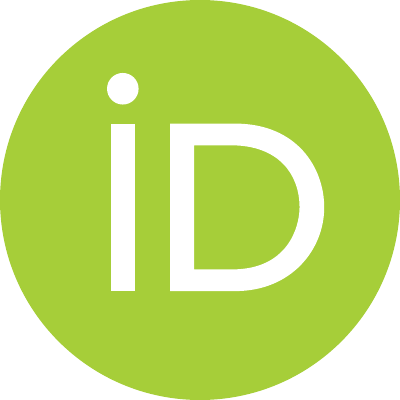}}}
\address{Nordita, KTH Royal Institute of Technology and Stockholm University,
Hannes Alfv\'ens vag 12, 106 91 Stockholm, Sweden}
\begin{abstract}
Preserving the precision of the parameter of interest in the presence
of environmental decoherence is an important yet challenging task
in dissipative quantum sensing. In this work, we study quantum metrology
when the decoherence effect is unraveled by a set of quantum measurements,
dubbed quantum measurement encoding. In our case, the estimation parameter
is encoded into a quantum state through a quantum measurement, unlike
the parameter encoding through a unitary channel in the decoherence-free
case or trace-preserving quantum channels in the case of decoherence.
We identify conditions for a precision-preserving measurement encoding.
These conditions can be employed to transfer metrological information
from one subsystem to another through quantum measurements. Furthermore,
postselected non-Hermitian sensing can also be viewed as quantum sensing
with measurement encoding. When the precision-preserving conditions
are violated in non-Hermitian sensing, we derive a universal formula
for the loss of precision. 
\end{abstract}
\maketitle
Quantum metrology and sensing is the subject of reducing noise in
experimental observations~\citep{sidhu2020geometric,liu2019quantum,pezz`e2018quantum,braun2018quantumenhanced,degen2017quantum,holevo2011probabilistic,helstrom1976quantum}
by utilizing quantum resources, such as coherences~\citep{giovannetti2006quantum},
entanglement~\citep{giovannetti2011advances,demkowicz-dobrzanski2014usingentanglement}
or many-body interactions~\citep{boixo2007generalized,roy2008exponentially,beau2017nonlinear,niezgoda2021manybody,yang2022variational,shi2024universal}.
However, environmental decoherence can kill all these advantages and
the precision can deteriorate significantly compared to the decoherence-free
scenario~\citep{fujiwara2008afibre,escher2011general,escher2012quantum,chin2012quantum,demkowicz-dobrzanski2012theelusive,alipour2014quantum,demkowicz-dobrzanski2014usingentanglement,smirne2016ultimate,beau2017nonlinear,demkowicz-dobrzanski2017adaptive,haase2018precision,bai2023floquet,liu2017quantum,yang2019memoryeffects,altherr2021quantum}.

\begin{figure}
\begin{centering}
\includegraphics[scale=0.35]{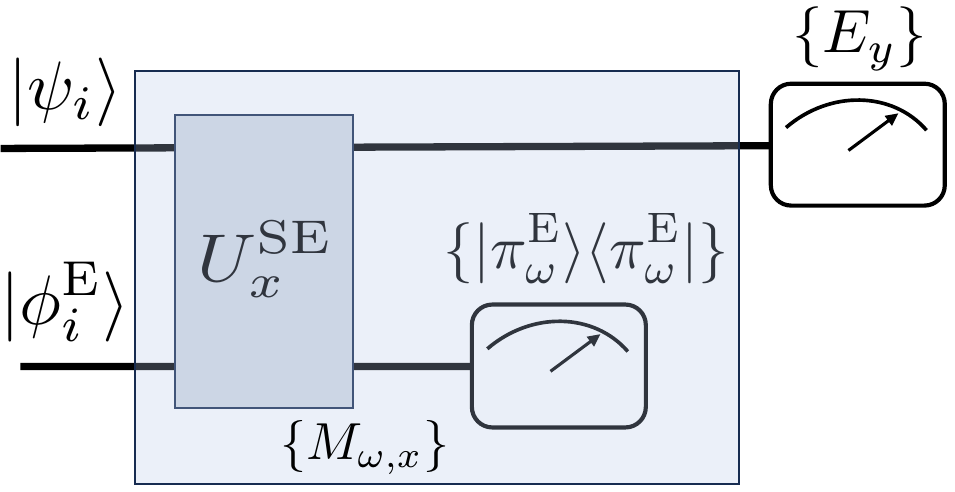}
\par\end{centering}
\caption{\label{fig:protocol}The protocol of quantum sensing with quantum
measurement encoding. The encoding measurement $M_{\omega,x}$, can
be effectively implemented by $U_{x}^{\text{SE}}$ and a projective
measurement $\{\ket{\pi_{\omega}^{\text{E}}}\bra{\pi_{\omega}^{\text{E}}}\}$,
encoding the estimation parameter $x$ into the state of the quantum
system. Here $M_{\omega,\,x}\!\equiv\!\braket{\pi_{\omega}^{\text{E}}\big|U_{x}^{\text{SE}}\big|\phi_{i}^{\text{E}}}$.
The POVM measurements $\{E_{y}\}$ acting on the system extracts the
metrological information into the measurement statistics. }
\end{figure}

On the other hand, since any trace-preserving decoherence channel
can be unravelled via a set of quantum measurements with all the measurement
outcomes discarded. In this work, we propose a framework for quantum
sensing with measurement encoding, where the measurement outcomes
and the post-measurement states are entirely or partially retained,
as shown in Fig.~\ref{fig:protocol}. While the protocol of measurement
encoding is related to postselected quantum metrology~\citep{yang2024theoryof,arvidsson-shukur2020quantum,lupu-gladstein2022negative,jenne2022unbounded,jordan2014technical,dressel2014colloquium,aharonov1988howthe},
it does not necessarily involve a postselection process that discards
the measurement outcomes or states and its precision lies in between
the decoherence-free and fully decohered protocols. As such, we regard
it as a fundamental metrological protocol. 

According to Uhlmann's theorem~\citep{uhlmann1976thetransition,nielsen2010quantum},
there exists an optimal environment under which the precision does
not deteriorate~\citep{fujiwara2008afibre,escher2011general,escher2012quantum}.
However, the structure of the optimal environment is largely uncharted,
which prevents such an observation from being of wide practical use.
Here, we show that for a dissipative trace-preserving quantum channel
to preserve the metrological information, there must exist a precision-preserving
encoding measurement that unravels the channel. 

For further applications, we show that it can be employed to perform
the transduction of quantum metrological information. Furthermore,
postselected non-Hermitian sensing~\citep{naghiloo2019quantum,ashida2020nonhermitian,clerk2022introduction,moiseyev2011nonhermitian,bender1998realspectra,hatano1996localization},
where only non-jump evolution is selected in the dissipative Markovian
GKSL (Gorini--Kossakowski--Sudarshan--Lindblad) master equation~\citep{lindblad1976onthe,gorini1976completely},
can be viewed as special case, where the encoding measurements are
concatenated in time. Quantum sensing with non-Hermitian physics has
attracted a lot of attention~\citep{wiersig2014enhancing,wiersig2016sensors,chen2017exceptional,hodaei2017enhanced,lau2018fundamental,zhang2019quantum,chu2020quantum,mcdonald2020exponentiallyenhanced,wiersig2020reviewof}
as well as debate~\citep{langbein2018noexceptional,ding2021experimental}.
We show that when the non-Hermitian dynamics is engineered through
postselection, it generically leads to loss of precision compared
to the corresponding decoherence-free scenario. We derive a universal
formula to quantify such a loss based on quantum collisional model~\citep{ciccarello2022quantum}.

\textit{Refined measurement and convexity inequalities}.\textit{---
}Throughout this work, operators with superscripts ``SE'' or ``E''
indicate they act on the system and environment jointly or only on
the environment, respectively. Operators with no superscripts indicate
that they act only on the primary system. As shown in Fig.~\ref{fig:protocol},
we consider an initial parameter-independent pure state of the system
$\varrho_{i}\!=\!\ket{\psi_{i}}\bra{\psi_{i}}$ that undergoes a generic
parameter-dependent quantum measurement channel described by $\{M_{\omega,\,x}\}_{\omega\in\Omega}$~\citep{kraus1971general,nielsen2010quantum},
where $\Omega$ denotes the set of all measurement outcomes. According
to Naimark Theorem~\citep{nielsen2010quantum,wilde2013quantum},
such a parameter-dependent measurement can be effectively realized
by unitary encoding that acts on both the system and the environment
and followed by a projective measurement on the environment, as shown
in Fig.~\ref{fig:protocol}. Two scenarios were compared in the previous
literature~\citep{escher2011general,escher2012quantum,chin2012quantum,demkowicz-dobrzanski2012theelusive,alipour2014quantum,demkowicz-dobrzanski2014usingentanglement,smirne2016ultimate,lu2010quantum,beau2017nonlinear,demkowicz-dobrzanski2017adaptive,haase2018precision,bai2023floquet,liu2017quantum,yang2019memoryeffects,altherr2021quantum}:(i)The
decoherence-free scenario---The experimenter has access to both the
system and the environment so that they can be regarded as a whole
closed system. The joint state is described by $\ket{\Psi_{x}^{\text{SE}}}\!\equiv\!U_{x}^{\text{SE}}\ket{\psi_{i}}\otimes\ket{\phi_{i}^{\text{E}}}$.
(ii) Decoherent scenario---Only the system's degree of freedom is
accessible. Then the decoherence effects induced by the environment
are characterized by a set of Kraus operators $\{M_{\omega,\,x}\}_{\omega\in\Omega}$.
The state after decoherence becomes $\rho_{x}\!=\!\sum_{\omega\in\Omega}M_{\omega,\,x}\varrho_{i}M_{\omega,\,x}^{\dagger}\!=\!\sum_{\omega\in\Omega}p(\omega|x)\sigma_{x|\omega}$,
where $p(\omega|x)\!=\!\text{Tr}(M_{\omega,\,x}^{\dagger}M_{\omega,\,x}\varrho_{i})$,
$\sigma_{x|\omega}\!=\!\tilde{\sigma}_{x|\omega}/p(\omega|x)$, $\tilde{\sigma}_{x|\omega}\!\equiv\!M_{\omega,\,x}\varrho_{i}M_{\omega,\,x}^{\dagger}$.
Here we consider the third scenario (iii) The measurement encoding
scenario, shown in Fig~(\ref{fig:protocol}). In addition, one may
further post-select the measurement outcomes to a specific set denoted
as $\checkmark$. 

We compare all three protocols under the same quantum resources, i.e.,
$U_{x}^{\text{SE}}$ is the same. The QFIs associated with each scenarios
are denoted as $I(\ket{\Psi_{x}^{\text{SE}}})$, $I(\rho_{x})$ and
$I(\sigma_{x|\omega})$. After the quantum measurement specified by
$\{M_{\omega,\,x}\}$, the state becomes $\sigma_{x}^{\text{SE}}\!=\!\sum_{\omega\in\Omega}p(\omega|x)\sigma_{x|\omega}\otimes\ket{\pi_{\omega}^{\text{E}}}\bra{\pi_{\omega}^{\text{E}}}$.
The QFI associated with $\sigma_{x}^{\text{SE}}$ is $I(\sigma_{x}^{\text{SE}})\!=\!\sum_{\omega\in\Omega}I_{\omega}(\sigma_{x}^{\text{SE}})$,
where $I_{\omega}(\sigma_{x}^{\text{SE}})\!\equiv\!p(\omega|x)I(\sigma_{x|\omega})\!+\!I_{\omega}^{\text{cl}}\left(p(\omega|x)\right)$
and $I_{\text{\ensuremath{\omega}}}^{\text{cl}}\left(p(\omega\big|x)\right)\!\equiv\!\left[\partial_{x}p(\omega\big|x)\right]^{2}/p(\omega\big|x)$~\citep{combes2014quantum,yang2024theoryof}.
We note that while the measurements involved in (iii) are separable
in the system and environment, (i) allows a wide joint collective
measurement. This naturally leads to the inequality,
\begin{equation}
I^{Q}\equiv I(\ket{\Psi_{x}^{\text{SE}}})\ge I(\sigma_{x}^{\text{SE}}).\label{eq:Psi-sigma-ineq}
\end{equation}
Alternatively, since $\sigma_{x}^{\text{SE}}$ is connected to $\ket{\Psi_{x}^{\text{SE}}}$
through the parameter-independent projective measurement $\{\ket{\pi_{\omega}^{\text{E}}}\}$,
intuitively the QFI cannot increase under any dissipative operations.
A rigorous justification of Eq.~(\ref{eq:Psi-sigma-ineq}) can be
found in Ref.~\citep{yang2024theoryof}. 

As mentioned before, discarding the measurement outcomes in (iii)
reduces to scenario (ii). As a result, intuitively we obtain,
\begin{equation}
I(\sigma_{x}^{\text{SE}})\ge I(\rho_{x}).\label{eq:sigma-rho-ineq}
\end{equation}
More rigorously, Eq.~(\ref{eq:sigma-rho-ineq}) is dictated by the
convexity of the QFI~\citep{alipour2015extended,pezze2014quantum,yu2013quantum}.
A few comments in order: Combing Eq.~(\ref{eq:Psi-sigma-ineq}) and
Eq.~(\ref{eq:sigma-rho-ineq}), we recover the Uhlmann inequality~\citep{uhlmann1976thetransition,nielsen2010quantum},
$I(\ket{\Psi_{x}^{\text{SE}}})\!\ge\!I(\rho_{x}).$ Ref.~\citep{escher2011general,escher2012quantum}
employs this inequality to derive the bounds for dissipative sensing
by optimizing over all possible Kraus operators. Nevertheless, saturating
the Uhlmann inequality is directly very challenging. As manifested
from the above expression for $I(\sigma_{x}^{\text{SE}})$, further
post-selecting the states $\sigma_{x|\omega}$ cannot increase the
QFI, i.e., $I(\sigma_{x}^{\text{SE}})\!\ge\!\sum_{\omega\in\checkmark}p(\omega|x)I(\sigma_{x|\omega})$.
Combining with inequality~(\ref{eq:Psi-sigma-ineq}), we obtain
\begin{equation}
I^{Q}\ge\sum_{\omega\in\checkmark}p(\omega|x)I(\sigma_{x|\omega}),\label{eq:PS-ineq}
\end{equation}
which clearly indicates that the average QFI over the post-selection
probability cannot outperform the one of post-selection free unitary
encoding. Ref.~\citep{ding2023fundamental} analyzed the average
precision for non-Hermitian sensing through error propagation formula.
Here, we would like to emphasize that their claim naturally follows
from the postselection inequality~(\ref{eq:PS-ineq}), which holds
for generic post-selection measurement encoding even beyond post-selected
non-Hermitian sensing.

One can further ask several fundamental questions regarding inequalities~(\ref{eq:Psi-sigma-ineq}-(\ref{eq:PS-ineq})),
which can lead to interesting yet uncharted physical consequences.
The saturation of (\ref{eq:sigma-rho-ineq}) and (\ref{eq:PS-ineq})
indicates the existence of dissipative quantum channels that preserves
the metrological information. The first step towards this goal is
to identify the conditions for precision-preserving encoding measurements,
which saturates Eq.~(\ref{eq:PS-ineq}). To this end, we consider
a refined version of the measurement inequality~(\ref{eq:Psi-sigma-ineq})~\citep{SM},
\begin{equation}
I_{\omega}(\ket{\Psi_{x}^{\text{SE}}})\ge I_{\omega}(\sigma_{x}^{\text{SE}}),\,\forall\omega,\label{eq:Refined-Meas-Ineq}
\end{equation}
where $I_{\omega}(\ket{\Psi_{x}^{\text{SE}}})$ and $I_{\omega}(\sigma_{x}^{\text{SE}})$
physically denote the QFIs associated with the particular measurement
operator $M_{\omega,\,x}$ before and after post-selection, respectively.
More specifically, $I_{\omega}(\ket{\Psi_{x}^{\text{SE}}})\!\equiv\!\braket{\partial_{x}^{\perp}\Psi_{x}^{\text{SE}}|\left(\mathbb{I}\otimes\ket{\pi_{\omega}^{\text{E}}}\bra{\pi_{\omega}^{\text{E}}}\right)\partial_{x}^{\perp}\Psi_{x}^{\text{SE}}}$
and $\ket{\partial_{x}^{\perp}\Psi_{x}^{\text{}}}$ is defined as~\citep{braunstein1994statistical,yang2019optimal,zhou2020saturating,liu2023optimal}
$\ket{\partial_{x}^{\perp}\Psi_{x}^{\text{SE}}}\!\equiv\!\ket{\partial_{x}\Psi_{x}^{\text{SE}}}\!-\!\ket{\Psi_{x}^{\text{SE}}}\braket{\Psi_{x}\big|\partial_{x}\Psi_{x}^{\text{SE}}}.$
Summing over all $\omega$ in Eq.~(\ref{eq:Refined-Meas-Ineq}) leads
to Eq.~(\ref{eq:Psi-sigma-ineq}). Similarly, a refined convexity
inequality can be also established~\citep{SM}$J_{\mu}(\sigma_{x}^{\text{SE}})\ge J_{\mu}(\rho_{x}),$
where $J_{\mu}(\sigma_{x}^{\text{SE}})\!\equiv\!\text{Tr}\left(\sum_{\omega}\tilde{\sigma}_{x|\omega}\tilde{L}_{x|\omega}E_{\mu}\tilde{L}_{x|\omega}\right)$,
$J_{\mu}(\rho_{x})\!\equiv\!\text{Tr}(\rho_{x}L_{x}E_{\mu}L_{x})$,
and $\{E_{\mu}\}$ is the set of positive operator-value measure (POVM)
operators acting on the system. $L_{x}$ and $\tilde{L}_{x|\omega}$
are the symmetric logarithmic derivative operator~\citep{paris2009quantum}
associated with the $\rho_{x}$ and $\tilde{\sigma}_{x|\omega}$,
respectively, defined as $\partial_{x}\rho_{x}\!\equiv\!(L_{x}\rho_{x}+\rho_{x}L_{x})/2$
and $\partial_{x}\tilde{\sigma}_{x|\omega}\!\equiv\!\left(\tilde{L}_{x|\omega}\tilde{\sigma}_{x|\omega}+\tilde{\sigma}_{x|\omega}\tilde{L}_{x|\omega}\right)/2$.
In particular, $\tilde{L}_{x|\omega}\!=\!\partial_{x}\ln p(\omega|x)+2\partial_{x}\sigma_{x|\omega}$.
Again, summing over all $\mu$ leads to Eq.~(\ref{eq:sigma-rho-ineq}).

\textit{Precision-preserving encoding measurements.}--- While the
saturation of the refined convexity is left for the future, we shall
focus on the saturation of the measurement inequality~(\ref{eq:Refined-Meas-Ineq}),
which leads to the precision-preserving encoding measurements. It
is straightforward to check that under the $U(1)$ parameter-dependent
gauge transformation $U_{x}^{\text{SE}}\to e^{\text{i}\theta_{x}}U_{x}^{\text{SE}}$,
$I(\ket{\Psi_{x}^{\text{SE}}})$, $I(\sigma_{x}^{\text{SE}})$ and
$I(\rho_{x})$ do not change. Such a gauge transformation leads to
$M_{\omega,\,x}\to e^{\text{i}\theta_{x}}M_{\omega,\,x}$. For fixed
initial state $\ket{\psi_{i}}$ and $\ket{\phi_{i}^{\text{E}}}$,
we can alway choose to work in the \textit{perpendicular} gauge such
that $\langle\!\langle\partial_{x}U_{x}^{\dagger\text{SE}}U_{x}^{\text{SE}}\rangle\!\rangle=\langle\!\langle U_{x}^{\dagger\text{SE}}\partial_{x}U_{x}^{\text{SE}}\rangle\!\rangle=0$,
where $\langle\!\langle\bullet\rangle\!\rangle$ denotes the average
over the joint system-environment state $\ket{\psi_{i}}\otimes\ket{\phi_{i}^{\text{E}}}$.
In this gauge, $\ket{\partial_{x}\Psi_{x}^{\text{SE}}}=\ket{\partial_{x}^{\perp}\Psi_{x}^{\text{SE}}}$
and the saturation condition of Eq.~(\ref{eq:Refined-Meas-Ineq})
becomes simple, which is our first main result~\citep{SM}:
\begin{thm}
\label{thm:lossless}\textup{In the perpendicular gauge, encoding
measurements are precision-preserving compared to scenario (i) if
and only if 
\begin{equation}
\langle\tilde{\psi}_{x|\omega}\big|\partial_{x}\tilde{\psi}_{x|\omega}\rangle\text{ is a real},\forall\omega\in\Omega.\label{eq:loss-encoding}
\end{equation}
Specifically, if postselection is involved, Eq.~(\ref{eq:PS-ineq})
is saturated if
\begin{align}
\langle\tilde{\psi}_{x|\omega}\big|\partial_{x}\tilde{\psi}_{x|\omega}\rangle & =0,\omega\in\checkmark,\label{eq:dMM-MdM}\\
\ket{\partial_{x}\tilde{\psi}_{x|\omega}} & =0,\,\omega\in\times,\label{eq:dM-cross}
\end{align}
where $\ket{\tilde{\psi}_{x|\omega}}\equiv M_{\omega,\,x}\ket{\psi_{i}}$.
In addition, for all $\omega\in\checkmark$, if $\ket{\partial_{x}\tilde{\psi}_{x|\omega}}\neq0$
then $p(\omega|x)=\braket{\tilde{\psi}_{x|\omega}\big|\tilde{\psi}_{x|\omega}}$
must be strictly positive.}
\end{thm}
From this theorem, we immediately know that a QFI-preserving dissipative
decoherence channel must allow quantum measurement unraveling such
that Eq.~(\ref{eq:loss-encoding}) is satisfied. Secondly, when Eq.~(\ref{eq:dMM-MdM})
does not exactly but approximately vanishes, it includes two possibilities:
The norm of $\ket{\tilde{\psi}_{x|\omega}}$ or $\ket{\partial_{x}\tilde{\psi}_{x|\omega}}$
is small or $\ket{\tilde{\psi}_{x|\omega}}$ and $\ket{\partial_{x}\tilde{\psi}_{x|\omega}}$
are almost orthogonal to each other up to a small error. We emphasize
here that only the latter case leads to a small loss of the QFI. Finally,
Theorem~\ref{thm:lossless} can be extended to the general gauge~\citep{SM}.
It is remarkable to note that Eq.~(\ref{eq:loss-encoding}) is gauge
invariant. 

Regardless of lossless or not, in the general gauge
\begin{equation}
I^{Q}\!=\!4(\langle G_{\Omega,\,x}\rangle-\langle F_{\Omega,\,x}\rangle),\label{eq:IQ-general}
\end{equation}
where $G_{\omega,\,x}\!\equiv\!\partial_{x}M_{\omega,\,x}^{\dagger}\partial_{x}M_{\omega,\,x}$,
$F_{\omega,\,x}\!\equiv\!\text{i}\partial_{x}M_{\omega,\,x}^{\dagger}M_{\omega,\,x}$,
$O_{\Omega,x}\equiv\sum_{\omega\in\Omega}O_{\omega,x}$ and $\langle\bullet\rangle$
denotes the expectation averaged over the system's initial state $\ket{\psi_{i}}$~\citep{escher2011general,SM}.
In the perpendicular gauge, $I^{Q}$ takes the simple form $I^{Q}\!=4\langle G_{\Omega,\,x}\rangle$
since $\langle F_{\Omega,\,x}\rangle=0$. We define the loss of measurement
encoding as $\kappa\equiv1-\sum_{\omega\in\checkmark}I_{\omega}(\sigma_{x}^{\text{SE}})/I^{Q}\in[0,\,1]$.
When only Eq.~(\ref{eq:dMM-MdM}) is satisfied, the average QFI over
the post-selection probability is $\sum_{\omega\in\checkmark}p(\omega|x)I(\sigma_{x|\omega})=4\langle G_{\checkmark,\,x}\rangle$
and $\kappa=\langle G_{\times,\,x}\rangle/\langle G_{\Omega,\,x}\rangle$
where $O_{\checkmark,x}\equiv\sum_{\omega\in\checkmark}O_{\omega,x}$
and $O_{\times,\,x}$ has a similar definition. 

\textit{Quantum measurements for metrological information transduction.}---
Previous works on post-selected metrology can be viewed as a special
case of precision-preserving measurement encoding, where certain post-measurement
states are discarded. Further more it concerns with either a single
quantum system~\citep{yang2024theoryof,lupu-gladstein2022negative,arvidsson-shukur2020quantum}
or the estimation of the interaction strength between two subsystems
as in the case of weak-value amplification~\citep{yang2024theoryof,dressel2014colloquium}. 

When the estimation parameter is initially sensed in one ancillary
system (the environment in Fig.~\ref{fig:protocol}), within the
coherence time of the ancilla, the metrological information must be
transferred to the target system (the primary system in Fig.~\ref{fig:protocol}).
We show that this goal can be achieved via a precision-preserving
measurement, which does not necessarily involve postselection. Since
$\{\ket{\pi_{\omega}^{\text{E}}}\bra{\pi_{\omega}^{\text{E}}}\}$
is a set of rank$-1$ projective operators, no information is left
in the environment afterwards. Furthermore, if Eq.~(\ref{eq:dMM-MdM})
in Theorem~\ref{thm:lossless} is satisfied for all $\omega\in\Omega$,
the system must contain all the metrological information. 

\begin{figure}
\begin{centering}
\includegraphics[scale=0.25]{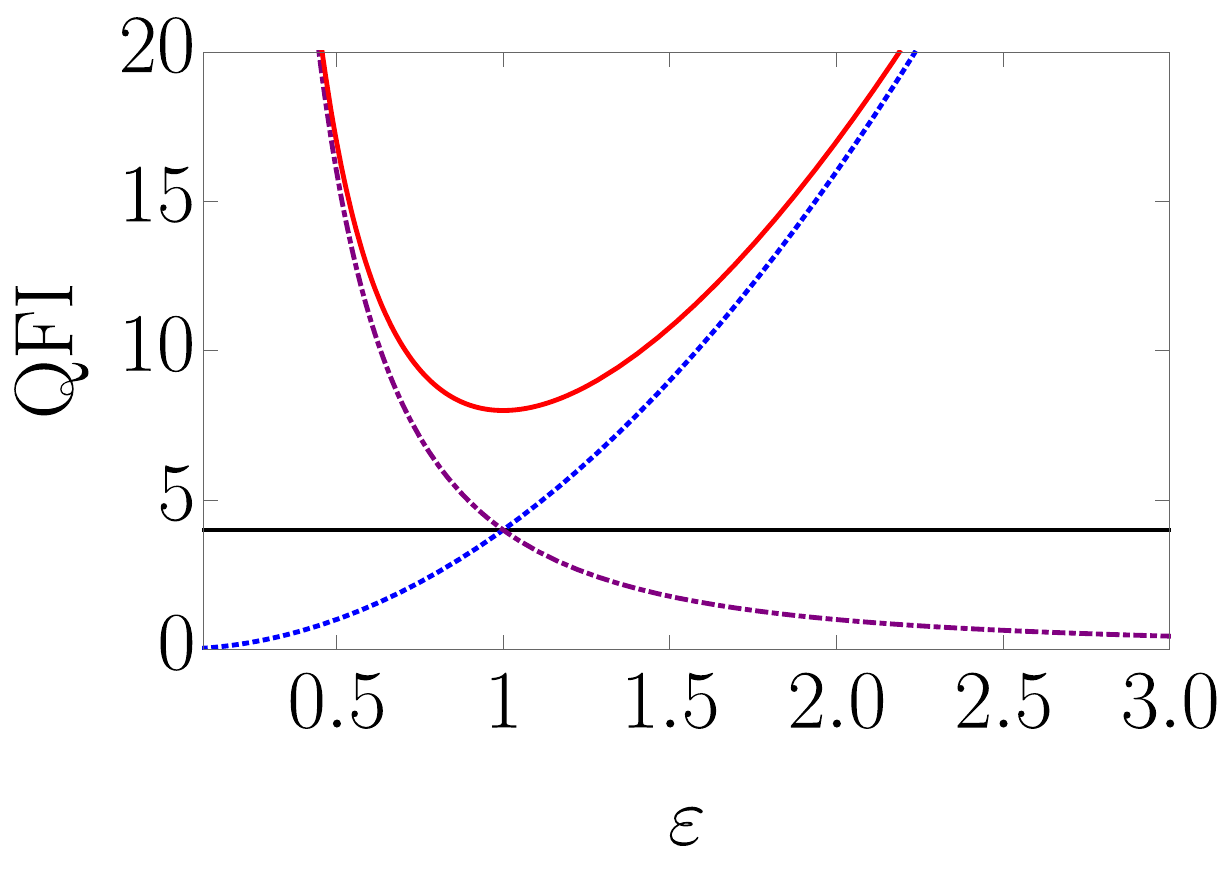}
\par\end{centering}
\caption{\label{fig:tranduction}Numerical calculation of various QFIs in the
two-qubit example in the main text. Values of parameters: $x=10^{-5}$
and sensing time $T=1$. The blue dashed and the purple dotted-dashed
lines correspond to the QFI $I(\sigma_{x|\omega=1})$ and $I(\sigma_{x|\omega=2})$,
respectively. The red solid line is the total sum of both. Similar
with weak value application, the QFI conditioned on a particular measurement
outcome can be way more larger than $I^{Q}$, but this is counterbalanced
by the probability of the outcome. In this case, the total average
QFI, i.e. $\sum_{\omega\in\Omega}p(\omega|x)I(\sigma_{x|\omega})$,
corresponding to the black solid line, equals to $I(\ket{\Psi_{x}^{\text{SE}}})=4$,
indicating the encoding measurement is precision-preserving.}
\end{figure}

We consider the following generic protocol: The ancillary system first
evolves under the Hamiltonian $H^{\text{E}}\!=\!xH_{0}^{\text{E}}$
for some $T$, followed by an entangling gate $U^{\text{I}}$. Then
a rank-$1$ projective measurement formed by the basis $\{\ket{\pi_{\omega}^{\text{E}}}\}$
is performed such that all the metrological information is transferred
to the target system. Throughout the process, we freeze the system's
free evolution, i.e., $H_{\text{S}}\!=\!0$ for simplicity. It can
readily calculated that $U_{x}^{\text{SE}}\!=\!U^{\text{I}}e^{-\text{i}xH_{0}^{\text{E}}T}$,
$U_{x}^{\dagger\text{SE}}\partial_{x}U_{x}^{\text{SE}}\!=\!-\text{i}H_{0}^{\text{E}}T$
and $I^{Q}\!=\!4T^{2}\text{Var}(H_{0}^{\text{E}})_{\ket{\phi_{i}^{\text{E}}}}$.
One can always shift $H_{0}^{\text{E}}$ by a constant such that $\braket{\phi_{i}^{\text{E}}\big|H_{0}^{\text{E}}\big|\phi_{i}^{\text{E}}}\!=\!0$
to move in the perpendicular gauge. Furthermore, with prior knowledge
of the estimation parameter and adaptive control~\citep{demkowicz-dobrzanski2014usingentanglement,yuan2015optimal,pang2017optimal,yang2017quantum,demkowicz-dobrzanski2017adaptive,yang2022variational},
we can assume $x$ is a very weak signal without loss of generality.
Then Eq.~(\ref{eq:dMM-MdM}) leads to 
\begin{equation}
\langle\psi_{i}|\otimes\langle\phi_{i}^{\text{E}}|U^{\text{I}\dagger}(\mathbb{I}\otimes\ket{\pi_{\omega}^{\text{E}}}\bra{\pi_{\omega}^{\text{E}}})U^{\text{I}}\ket{\psi_{i}}\otimes\ket{\phi_{i}^{\text{E}\perp}}=0,\label{eq:Optimal-Encoding}
\end{equation}
where we have used the weak signal approximation to set $U_{x}^{\text{SE}}\!=\!U_{\text{}}^{\text{I}}$
and $\ket{\phi_{i}^{\text{E}\perp}}\!\equiv\!H_{0}^{\text{E}}\ket{\phi_{i}^{\text{E}}}/\sqrt{\text{Var}(H_{0}^{\text{E}})_{\ket{\phi_{i}^{\text{E}}}}}$
is orthogonal to $\ket{\phi_{i}^{\text{E}}}$ in the perpendicular
gauge. Intuitively, one can construct the following control-$U$ gate,
\begin{equation}
U^{\text{I}}=\mathbb{I}\otimes\left(\mathbb{I}^{\text{E}}-\ket{\phi_{i}^{\text{E}\perp}}\bra{\phi_{i}^{\text{E}\perp}}\right)+\mathbb{U}\otimes\ket{\phi_{i}^{\text{E}\perp}}\bra{\phi_{i}^{\text{E}\perp}},\label{eq:UI-expression}
\end{equation}
where $\mathbb{U}\ket{\psi_{i}}$ is orthogonal to $\ket{\psi_{i}}$.
Then Eq.~(\ref{eq:Optimal-Encoding}) becomes $\langle\psi_{i}|\otimes\langle\phi_{i}^{\text{E}}|\mathbb{I}\otimes\ket{\pi_{\omega}^{\text{E}}}\langle\pi_{\omega}^{\text{E}}\ket{\psi_{i}^{\perp}}\otimes\ket{\phi_{i}^{\text{E}\perp}}\!=\!0$,
which can satisfies by any rank-$1$ projective measurements since
$\braket{\psi_{i}|\psi_{i}^{\perp}}\!=\!0$. The simplest choice one
can construct is the following:
\begin{equation}
\ket{\pi_{1}^{\text{E}}(\varepsilon)}=\frac{\ket{\phi_{i}^{\text{E}}}+\varepsilon\ket{\phi_{i}^{\text{E}\perp}}}{\sqrt{1+\varepsilon^{2}}},\,\ket{\pi_{2}^{\text{E}}(\varepsilon)}=\frac{\ket{\phi_{i}^{\text{E}\perp}}-\varepsilon\ket{\phi_{i}^{\text{E}}}}{\sqrt{1+\varepsilon^{2}}},
\end{equation}
where $\varepsilon$ is an arbitrary real number and other projectors
are formed by orthonormal basis outside the subspace spanned by $\ket{\phi_{i}^{\text{E}}}$
and $\ket{\phi_{i}^{\text{E}\perp}}$. Since $U_{x}^{\text{SE}}\!\approx\!U^{\text{I}}$
and $\partial_{x}U_{x}^{\text{SE}}\!=\!-\text{i}TU^{\text{I}}H_{0}^{\text{E}}$
, we obtain $M_{\omega}(\varepsilon)\!=\!\braket{\pi_{\omega}^{\text{E}}(\varepsilon)\big|U^{\text{I}}\big|\phi_{i}^{\text{E}}}$,
$\partial_{x}M_{\omega}(\varepsilon)\!=\!-\text{i}\sqrt{I^{Q}}\braket{\pi_{\omega}^{\text{E}}(\varepsilon)\big|U^{\text{I}}\big|\phi_{i}^{\text{E}\perp}}/2$.
Thus it is straightforward to show $M_{1}(\varepsilon)\!\approx\!-M_{2}(\varepsilon)/\varepsilon\!\approx\!\mathbb{I}/\sqrt{1+\varepsilon^{2}}$
and $\partial_{x}M_{1}(\varepsilon)/\varepsilon\!\approx\!\partial_{x}M_{2}(\varepsilon)\!\approx\!-\text{i}\sqrt{I^{Q}}\mathbb{U}/\left(2\sqrt{1+\varepsilon^{2}}\right)$.
Note that although the magnitude of these measurement operators can
be small, their sensitivity to change of the estimation parameter
can be large. 

For example, consider a minimum two-qubit model with $H^{\text{S}}\!=\!0$,
$H^{\text{E}}\!=\!x\sigma_{z}$, $\ket{\phi_{i}^{\text{E}}}\!=\!\ket{+x}$,
where $\ket{0}$ and $\ket{1}$ denotes the excited and ground states
of $\sigma_{z}$ and $\ket{\pm x}\!=\!(\ket{0}\!\pm\!\ket{1})/\sqrt{2}$.
The initial state of the system is $\ket{\psi_{i}}\!=\!\ket{0}$.
The controlled-$U$ gate then becomes $U^{\text{I}}\!=\!\mathbb{I}\otimes\ket{+x}\bra{+x}\!+\!\sigma_{x}\otimes\ket{-x}\bra{-x}$.
The plots of the various QFIs for post-measurement states for small
$x$ are shown in Fig. ~\ref{fig:tranduction}. In the perpendicular
gauge, the QFI for post-measurement states is the norm of $\ket{\partial_{x}\tilde{\psi}_{x|\omega}}$.
In the limit $\varepsilon\to0$ or $\varepsilon\to\infty$, one of
the post-measurement state can be discarded because it satisfies Eq.~(\ref{eq:dM-cross})
and thus carries no information, as shown in Fig. ~\ref{fig:tranduction}. 

\textit{Postselected non-Hermitian sensing}. --- Now we are in a
position to how non-Hermitian sensing is connected to the measurement
encoding problem above. Consider the non-Hermitian dynamics, 
\begin{equation}
\dot{\rho}_{x}(t)=-\text{i}\left(H_{x}^{\text{nh}}(t)\rho_{x}(t)-\rho_{x}(t)H_{x}^{\text{nh}\dagger}(t)\right),
\end{equation}
where $H_{1}(t)$ is some control Hamiltonian,
\begin{equation}
H_{x}^{\text{nh}}(t)\equiv H_{0x}(t)+H_{1}(t)-\frac{\text{i}}{2}\sum_{j}\gamma_{j}(t)L_{j}^{\dagger}L_{j},\label{eq:H-NH}
\end{equation}
and $\gamma_{j}(t)\ge0$. It is well known that the GKSL Markovian
master equation~\citep{gorini1976completely,lindblad1976onthe,breuer2007thetheory,rivas2011openquantum}
describing open-system dynamics can be unravelled by a sequel of measurements
in continuous time, which leads to stochastic evolution of the state~\citep{wiseman2014quantum,jacobs2014quantum}.
Non-Hermitian dynamics can be further effectively realized by eliminating
state trajectories with quantum jumps~\citep{naghiloo2019quantum}. 

\begin{figure}
\begin{centering}
\includegraphics[scale=0.25]{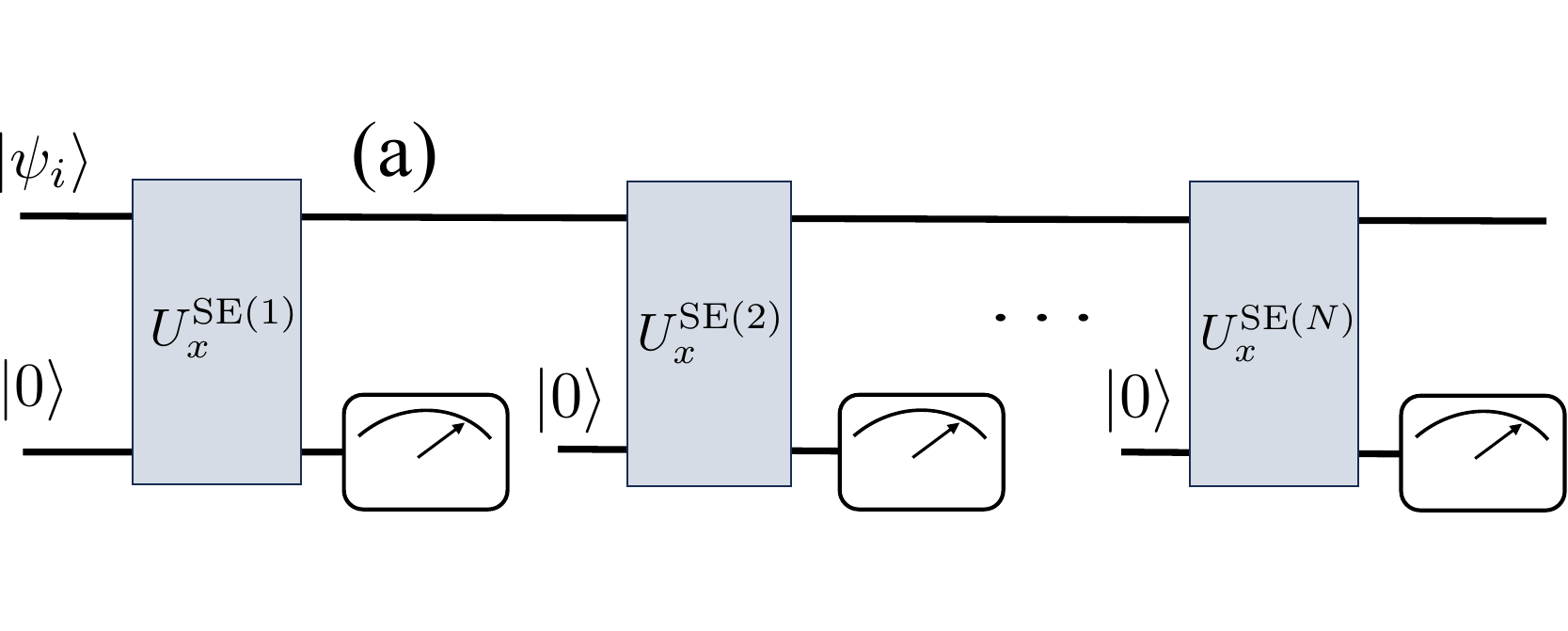}
\par\end{centering}
\begin{centering}
\includegraphics[scale=0.25]{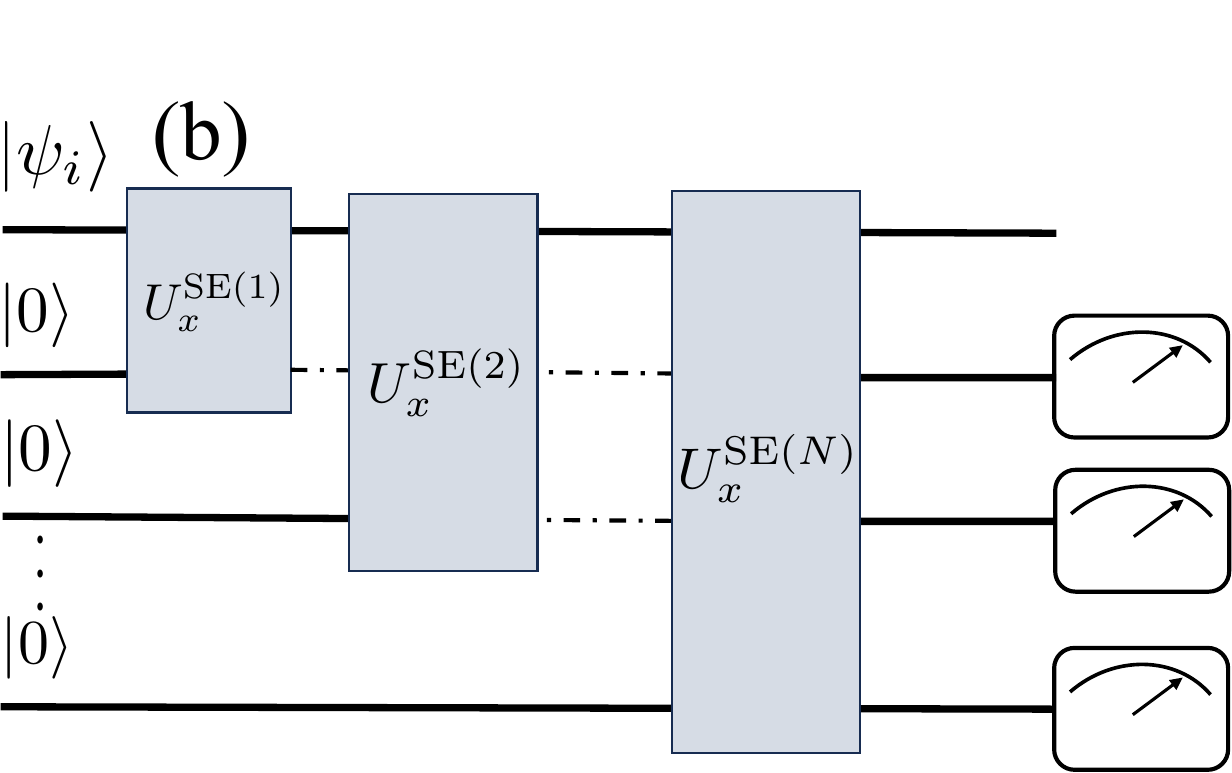}
\par\end{centering}
\caption{\label{fig:NH-Sensing}(a) Unravelling the discrete GKSL master equation
with the collision model(b) Deferred measurement principle for post-selected
non-Hermitian evolution. }

\end{figure}

To facilitate our subsequent discussion, we first consider the discrete
time evolution and take the continuum limit at the end. We denote
the probe time as $T$. Then discrete non-Hermitian evolution is given
by $M_{\checkmark,\,x}\!\equiv\!M_{\checkmark,\,x}^{(N)}\cdots M{}_{\checkmark,\,x}^{(2)}M_{\checkmark,\,x}^{(1)}$,
where $M_{\checkmark,\,x}^{(n)}\!=\!\mathbb{I}-\text{i}H_{x}^{\text{nh}}(t_{n})\Delta t$,
$M_{j}^{(n)}\!=\!\sqrt{\gamma_{j}(t_{n})\Delta t}L_{j}$, and $t_{n}\!=\!n\Delta t$.The
discrete-time analog of quantum trajectories correspond to the quantum
collision model~\citep{ciccarello2022quantum,rau1963relaxation,ziman2002diluting,cattaneo2021collision}
with a concatenation of measurements, as depicted in Fig.~\ref{fig:NH-Sensing}(a).
The discrete non-Hermitian evolution is implemented by post-selecting
the outcomes of these measurements. We define $\braket{\pi_{\checkmark}|U_{x}^{\text{SE}(n)}|0}\!\equiv\!M_{\checkmark,\,x}^{(n)}$,
$\,\braket{\pi_{j}|U_{x}^{\text{SE}(n)}|0}\!\equiv\!M_{j,\,x}^{(n)},\,$
and $U_{x}^{\text{SE}}\!\equiv\!U_{x}^{\text{SE}(N)}\cdots U_{x}^{\text{SE}(1)}$.
As the measurement operators satisfy the approximate completeness
relations $M_{\checkmark,\,x}^{(n)^{\dagger}}M_{\checkmark,\,x}^{(n)}\!+\!\sum_{j\in\times}M_{j,\,x}^{(n)\dagger}M_{j,\,x}^{(n)}\!=\!\mathbb{I}\!+\!O(\Delta t^{2})$
and $\sum_{n=1}^{N}\sum_{j}M_{jn,\,x}^{\dagger}M_{jn,\,x}\!+\!M_{\checkmark,\,x}^{\dagger}M_{\checkmark,\,x}\!=\!\mathbb{I}\!+\!O(N\Delta t^{2})$~\citep{SM},
so are the unitary operators, i.e., $U_{x}^{\text{SE}\dagger(n)}U_{x}^{\text{SE}(n)}\!=\!\mathbb{I}+O(\Delta t^{2})$
and $U_{x}^{\text{SE}\dagger}U_{x}^{\text{SE}}\!=\!\mathbb{I}+O(N\Delta t^{2})$. 

According to the deferred measurement principle, the post-selection
measurement can be always deferred to the very end, as shown in Fig.
~\ref{fig:NH-Sensing}(b). With this observation, we recognize that
the discrete non-Hermitian sensing belongs to the measurement encoding
problem with a single retained state. The retained measurement outcome
is $(\checkmark,\,\checkmark,\,\cdots\checkmark)$ corresponding to
the measurement operator $M_{\checkmark,\,x}$. All the remaining
outcomes lies in the discarded set $\times$, with the measurement
operators $M_{jn,\,x}\!\equiv\!M_{j}^{(n)}M_{\checkmark,\,x}^{(n-1)}\cdots M_{\checkmark,\,x}^{(1)}$,
and the environment initial state $\ket{\phi_{i}^{\text{E}}}=\ket{0}^{\otimes N}$.
We can then apply Theorem~\ref{thm:lossless} and take the continuous-time
limit $\Delta t\to0$, $N\to\infty$ $N\Delta t=T$ to identify QFI-preserving
conditions for post-selected non-Hermitian sensing. Note that in the
continuous-time limit, $M_{\checkmark,\,x}(T)\!\equiv\!\mathcal{T}e^{-\text{i}\int_{0}^{T}H_{x}^{\text{nh}}(\tau)d\tau}$
and the approximate completeness relation for the measurement operators
in the discrete case now becomes the exact integral completeness relation:
\begin{equation}
\int_{0}^{T}M_{\checkmark,\,x}^{\dagger}(t)L^{2}(t)M_{\checkmark}(t)dt+M_{\checkmark,\,x}^{\dagger}(T)M_{\checkmark,\,x}(T)=\mathbb{I},\label{eq:int-completeness}
\end{equation}
where $L^{2}(t)\equiv\sum_{j}\gamma_{j}(t)L_{j}^{\dagger}L_{j}$.
Eq.~(\ref{eq:int-completeness}) can be straightforwardly verified
by taking derivatives with respect to $T$ and then applying the non-Hermitian
dynamics $\text{i}M_{\checkmark,\,x}(t)=H_{\text{nh}}(t)M_{\checkmark,\,x}(t)$.
Next, we present our second main result:
\begin{thm}
\textup{\label{thm:Lossles-NH-sensing}Non-Hermitian sensing with
is lossless compared with the quantum collision model shown in Fig.~\ref{fig:NH-Sensing}(b)
in the continuous-time limit, if and only if (a) The post-selection
probability is insensitive to the estimation parameter, i.e., $\partial_{x}\langle E_{\checkmark,\,x}\rangle\!=\!0$,
where $E_{\checkmark,\,x}\!\equiv\!M_{\checkmark,\,x}^{\dagger}M_{\checkmark,\,x}$.
(b) For all $j$ and $t\in[0,\,T]$, $L_{j}\ket{\tilde{\xi}_{\checkmark,\,x}(t)}\!=\!0$,
where $\ket{\tilde{\xi}_{\checkmark,\,x}(t)}\!\equiv\![\ket{\partial_{x}\tilde{\psi}_{x|\checkmark}(t)}\!+\!\text{i}\partial_{x}\theta_{\checkmark,\,x}\ket{\tilde{\psi}_{x|\checkmark}(t)}]$
and $\partial_{x}\theta_{\checkmark,\,x}\!\equiv\!-\text{Re}\langle F_{\checkmark,\,x}\rangle/\langle E_{\checkmark,\,x}\rangle\!=\!\text{Im}\langle\partial_{x}M_{\checkmark,\,x}^{\dagger}M_{\checkmark,\,x}\rangle/\langle E_{\checkmark,\,x}\rangle$.}
\end{thm}
A few comments in order. One special case is that if $L_{j}\mathcal{T}e^{-\text{i}\int_{0}^{T}[H_{0x}(t)+H_{1}(t)]dt}\ket{\psi_{i}}\!=\!0$
for all $j$, which means that no jump occurs, then non-Hermitian
dynamics is effective unitary for the initial state $\ket{\psi_{i}}$.
In this case, $\langle E_{\checkmark,\,x}\rangle\!=\!0$ and $L_{j}\ket{\tilde{\psi}_{x|\checkmark}(t)}\!=\!0$,
which clearly satisfies Theorem~\ref{thm:Lossles-NH-sensing}. Secondly,
it can be calculated that~\citep{SM}:
\begin{align}
\langle G_{\Omega,\,x}\rangle & =\langle G_{\checkmark,\,x}\rangle+\int_{0}^{T}\langle\partial_{x}M_{\checkmark,\,x}^{\dagger}(t)L^{2}(t)\partial_{x}M_{\checkmark,\,x}(t)\rangle dt,\label{eq:G-Omg-nh}\\
\langle F_{\Omega,\,x}\rangle & =\langle F_{\checkmark,\,x}\rangle+\text{i}\int_{0}^{T}\langle\partial_{x}M_{\checkmark,\,x}^{\dagger}(t)L^{2}(t)M_{\checkmark,\,x}(t)\rangle dt.\label{eq:F-Omg-nh}
\end{align}

As a result, when the conditions in Theorem~\ref{thm:Lossles-NH-sensing}
is not satisfied or partially satisfied, one can estimate the loss
of the QFI compared with the unitary quantum collision model, i.e.,
$\kappa\!=\!\left[1\!-\!p(\checkmark|x)I(\sigma_{x|\checkmark})/I^{Q}\right]$,
where $p(\checkmark|x)I(\sigma_{x|\checkmark})\!=\!4\left[\langle G_{\checkmark,\,x}\rangle\!-\!\frac{|\langle F_{\checkmark,\,x}\rangle|^{2}}{\langle E_{\checkmark,\,x}\rangle}\right]$~\citep{yang2024theoryof}
and $I^{Q}$ can be calculated using Eqs.~(\ref{eq:IQ-general},~\ref{eq:G-Omg-nh},~\ref{eq:F-Omg-nh}).
This is our third main result. We would like to emphasize the universality
of the loss formula, which does not depend on the microscopic environment
and coupling details in the quantum collision model. 

To illustrate the application of the loss formula, we consider a simple
yet non-trivial case of pure dephasing channels with multiplicative
estimation Hamiltonian, where $H_{0x}=xH_{0}$ and $H_{0}$, $H_{1}(t)$,
$L$ mutually commute. Furthermore, we assume $\gamma_{j}(t)$ is
time-independent for simplicity. It can be readily calculated that
$M_{\checkmark,\,x}(T)\!=\!e^{-\text{i}[xH_{0}T+\int_{0}^{T}H_{1}(t)dt]}e^{-\frac{1}{2}L^{2}T}$,
$\langle E_{\checkmark,\,x}\rangle\!=\!\langle e^{-L^{2}T}\rangle$,
$\langle F_{\checkmark,\,x}\rangle\!=\!-T\langle e^{-L^{2}T}H_{0}\rangle$,
and $\langle G_{\checkmark,\,x}\rangle\!=\!T^{2}\langle e^{-L^{2}T}H_{0}^{2}\rangle$.
It is straightforward to check that the integral completeness relation~(\ref{eq:int-completeness})
is satisfied. Moreover, physically since the the system-environment
coupling commutes with $H_{0}$ and $H_{1}(t)$, it is expected that
QFI for closed-system must be $I^{Q}\!=\!4T^{2}\left(\langle H_{0}^{2}\rangle-\langle H_{0}\rangle^{2}\right)$,
which is the same as the prediction using Eqs.~(\ref{eq:IQ-general},~\ref{eq:G-Omg-nh},~\ref{eq:F-Omg-nh}).
These observations demonstrate the validity of the theory presented
here. Finally, the loss for the dephasing model is given by
\begin{equation}
\kappa=1-\frac{\langle e^{-L^{2}T}H_{0}^{2}\rangle-(\langle e^{-L^{2}T}H_{0}\rangle)^{2}/\langle e^{-L^{2}T}\rangle}{\langle H_{0}^{2}\rangle-\langle H_{0}\rangle^{2}}.
\end{equation}
In the case where $H_{0}=\sigma_{z}$, $\ket{\psi_{i}}=\ket{+x}$,
and $L_{1}=L_{1}^{\dagger}=\sigma_{z}$, we find $I(\sigma_{x|\checkmark})=I^{Q}=4T^{2}$
with loss $\kappa=1-e^{-\gamma_{1}T}$, which increases with the probe
time.

\textit{Conclusion. ---} We provide a framework for quantum metrology
using quantum measurement encoding, which may open a new research
line. Our results imply that precision-preserving measurement encoding
is necessary for dissipative sensing to preserve the metrological
information and find applications in the transduction of metrological
information and postselected non-Hermitian sensing. We hope our findings
here can inspire further studies on dissipative QFI-preserving quantum
channels. Future directions include investigating the interplay between
many-body and post-selection effects, generalizing to mixed states,
and exploring non-Markovian measurement encoding, etc.

\textit{Acknowledgement. ---} The author would like to thank Adolfo
del Campo, Luiz Davidovich and Hai-Long Shi for useful comments and
Hongzhe Zhou for hosting his visit to Tsung-Dao Lee Institute at Shanghai
Jiao Tong University, during which this work was completed. This work
was funded by the Wallenberg Initiative on Networks and Quantum Information
(WINQ) program. 

\let\oldaddcontentsline\addcontentsline     
\renewcommand{\addcontentsline}[3]{}         

\bibliography{Post-Selected-Many-body-Metrology}

\clearpage{}

\newpage{}

\setcounter{equation}{0} \setcounter{section}{0} \setcounter{subsection}{0}
\global\long\def\theequation{S\arabic{equation}}%
\onecolumngrid \setcounter{enumiv}{0} 

\setcounter{equation}{0} \setcounter{section}{0} \setcounter{subsection}{0} \renewcommand{\theequation}{S\arabic{equation}} \onecolumngrid \setcounter{enumiv}{0} \setcounter{figure}{0} \renewcommand{\thefigure}{S\arabic{figure}}
\begin{center}
\textbf{\large{}Supplemental Material}{\large\par}
\par\end{center}

In this Supplemental Material, we provide proof of the refined convexity
inequality of the QFI, proof Theorem~\ref{thm:lossless} and its
form in generic gauge, the details of the calculation for metrological
transduction using quantum measurements, the approximate completeness
relation, Proof of Theorem~\ref{thm:Lossles-NH-sensing}, and the
integral representations of $\langle G_{\Omega,x}\rangle$ and $\langle F_{\Omega,x}\rangle$
for the quantum collision model, which is useful in evaluating the
loss in post-selected non-Hermitian sensing.

\let\addcontentsline\oldaddcontentsline     

\tableofcontents{}

\addtocontents{toc}{\protect\thispagestyle{empty}}
\pagenumbering{gobble}

\section{Refined convexity inequality of the QFI}

The proof of the convexity property Eq.~(\ref{eq:sigma-rho-ineq})
can be found in Refs.~\citep{yu2013quantum,alipour2015extended,pezze2014quantum}.
Here we provide a refined convexity inequality. Let us note the classical
Fisher information associated with the POVM measurement $\{E_{\mu}\}$
on the state is
\begin{equation}
J_{\mu}^{\text{cl}}(\rho_{x};E_{\mu})=\frac{[\text{Tr}(\partial_{x}\rho_{x}E_{\mu})]^{2}}{\text{Tr}(\rho_{x}E_{\mu})}=\frac{[\sum_{\omega}\text{Tr}(\partial_{x}\tilde{\sigma}_{x|\omega}E_{\mu})]^{2}}{\text{Tr}(\rho_{x}E_{\mu})},
\end{equation}
where 
\begin{equation}
\tilde{\sigma}_{x|\omega}=p(\omega|x)\sigma_{x|\omega}.
\end{equation}
According to Ref.~\citep{yang2019optimal}, we know 
\begin{equation}
J_{\mu}^{\text{cl}}(\rho_{x};E_{\mu})\leq J_{\mu}(\rho_{x})\equiv\text{Tr}(\rho_{x}L_{x}E_{\mu}L_{x})=\sum_{\omega}\text{Tr}(\tilde{\sigma}_{x|\omega}L_{x}E_{\mu}L_{x}).\label{eq:qCRB}
\end{equation}
More specifically, one can show 
\begin{align}
J_{\mu}^{\text{cl}}(\rho_{x};E_{\mu}) & =\frac{\left[\text{Re}\text{Tr}(\rho_{x}E_{\mu}L_{x})\right]^{2}}{\text{Tr}(\rho_{x}E_{\mu})}=\frac{\left[\text{Re}\sum_{\omega}\text{Tr}\left(\tilde{\sigma}_{x|\omega}E_{\mu}L_{x}\right)\right]^{2}}{\text{Tr}(\rho_{x}E_{\mu})}\nonumber \\
 & \leq\frac{\left[\big|\sum_{\omega}\text{Tr}\left(\tilde{\sigma}_{x|\omega}E_{\mu}L_{x}\right)\big|\right]^{2}}{\text{Tr}(\rho_{x}E_{\mu})}\label{eq:Re-Ineq}\\
 & \leq\frac{\left[\sum_{\omega}\big|\text{Tr}\left(\tilde{\sigma}_{x|\omega}E_{\mu}L_{x}\right)\big|\right]^{2}}{\text{Tr}(\rho_{x}E_{\mu})}\label{eq:Tri-ineq}\\
 & \leq\frac{\left[\sum_{\omega}\sqrt{\text{Tr}\left(\tilde{\sigma}_{x|\omega}E_{\mu}\right)}\sqrt{\text{Tr}\left(\tilde{\sigma}_{x|\omega}L_{x}E_{\mu}L_{x}\right)}\right]^{2}}{\text{Tr}(\rho_{x}E_{\mu})}\label{eq:CS1}\\
 & \leq\frac{\sum_{\omega}\text{Tr}\left(\tilde{\sigma}_{x|\omega}E_{\mu}\right)}{\text{Tr}(\rho_{x}E_{\mu})}\sum_{\omega}\text{Tr}\left(\tilde{\sigma}_{x|\omega}L_{x}E_{\mu}L_{x}\right)\label{eq:CS2}\\
 & =J_{\mu}(\rho_{x}),\label{eq:qCRB-detail}
\end{align}
where in the inequality~(\ref{eq:Tri-ineq}) we have used the triangle
inequality and in equalities~(\ref{eq:CS1}) and (\ref{eq:CS2}),
we have used the Cauchy-Schwarz inequality. We define~\citep{alipour2015extended}
\begin{equation}
\partial_{x}\tilde{\sigma}_{x|\omega}=\frac{\tilde{L}_{x|\omega}\tilde{\sigma}_{x|\omega}+\tilde{\sigma}_{x|\omega}\tilde{L}_{x|\omega}}{2}.
\end{equation}
It is straightforward to verify that 
\begin{equation}
\tilde{L}_{x|\omega}=\partial_{x}\ln p(\omega|x)+L_{x|\omega},
\end{equation}
where $L_{x|\omega}$ is defined as 
\begin{equation}
\partial_{x}\sigma_{x|\omega}=\frac{L_{x|\omega}\sigma_{x|\omega}+\sigma_{x|\omega}L_{x|\omega}}{2}.
\end{equation}
Thus we find 
\begin{equation}
J_{\mu}^{\text{cl}}(\rho_{x};E_{\mu})=\frac{\left[\sum_{\omega}\text{Tr}(\partial_{x}\tilde{\sigma}_{x|\omega}E_{\mu})\right]^{2}}{\text{Tr}(\rho_{x}E_{\mu})}=\frac{\left[\text{Re}\sum_{\omega}\text{Tr}(\tilde{\sigma}_{x|\omega}E_{\mu}\tilde{L}_{x|\omega})\right]^{2}}{\text{Tr}(\rho_{x}E_{\mu})}.
\end{equation}
Following similar procedure in inequality~(\ref{eq:CS2}), it can
be readily shown that
\begin{align}
J_{\mu}^{\text{cl}}(\rho_{x};E_{\mu}) & \leq\frac{\left[\big|\sum_{\omega}\text{Tr}\left(\tilde{\sigma}_{x|\omega}E_{\mu}\tilde{L}_{x|\omega}\right)\big|\right]^{2}}{\text{Tr}(\rho_{x}E_{\mu})}\nonumber \\
 & \leq\frac{\left[\sum_{\omega}\big|\text{Tr}\left(\tilde{\sigma}_{x|\omega}E_{\mu}\tilde{L}_{x|\omega}\right)\big|\right]^{2}}{\text{Tr}(\rho_{x}E_{\mu})}\\
 & \leq\frac{\left[\sum_{\omega}\sqrt{\text{Tr}\left(\tilde{\sigma}_{x|\omega}E_{\mu}\right)}\sqrt{\text{Tr}\left(\tilde{\sigma}_{x|\omega}\tilde{L}_{x|\omega}E_{\mu}\tilde{L}_{x|\omega}\right)}\right]^{2}}{\text{Tr}(\rho_{x}E_{\mu})}\\
 & \leq\frac{\left[\sum_{\omega}\text{Tr}\left(\tilde{\sigma}_{x|\omega}E_{\mu}\right)\right]\left[\sum_{\omega}\text{Tr}\left(\tilde{\sigma}_{x|\omega}\tilde{L}_{x|\omega}E_{\mu}\tilde{L}_{x|\omega}\right)\right]}{\text{Tr}(\rho_{x}E_{\mu})}\\
 & =\sum_{\omega}\text{Tr}\left(\tilde{\sigma}_{x|\omega}\tilde{L}_{x|\omega}E_{\mu}\tilde{L}_{x|\omega}\right).
\end{align}
We define 
\begin{equation}
L_{x}^{\text{SE}}=\sum_{\omega}\tilde{L}_{x|\omega}\otimes\ket{\pi_{\omega}^{\text{E}}}\bra{\pi_{\omega}^{\text{E}}},
\end{equation}
which satisfies 
\begin{equation}
\partial_{x}\sigma_{x}^{\text{SE}}=\frac{L_{x}^{\text{SE}}\sigma_{x}^{\text{SE}}+\sigma_{x}^{\text{SE}}L_{x}^{\text{SE}}}{2}.
\end{equation}
Then it is straightforward to calculate that 
\begin{align}
\text{Tr}\left(\sum_{\omega}\tilde{\sigma}_{x|\omega}\tilde{L}_{x|\omega}E_{\mu}\tilde{L}_{x|\omega}\right) & =\text{Tr}_{\text{SE}}\left(\sum_{\omega}\tilde{\sigma}_{x|\omega}\otimes\ket{\pi_{\omega}^{\text{E}}}\bra{\pi_{\omega}^{\text{E}}}L_{x}^{\text{SE}}(E_{\mu}\otimes\mathbb{I}^{\text{E}})L_{x}^{\text{SE}}\right)\nonumber \\
 & =\text{Tr}_{\text{SE}}\left(\sigma_{x}^{\text{SE}}L_{x}^{\text{SE}}(E_{\mu}\otimes\mathbb{I}^{\text{E}})L_{x}^{\text{SE}}\right)\equiv J_{\mu}(\sigma_{x}^{\text{SE}}).
\end{align}
Thus, we find 
\begin{equation}
J_{\mu}^{\text{cl}}(\rho_{x};E_{\mu})\leq J_{\mu}(\sigma_{x}^{\text{SE}}).\label{eq:refined-ConIneq-Intermediate}
\end{equation}
Since $\sigma_{x|\omega}$ is a pure state, 
\begin{equation}
L_{x|\omega}=2\partial_{x}\sigma_{x|\omega},\,\tilde{L}_{x|\omega}=\partial_{x}\ln p(\omega|x)+2\partial_{x}\sigma_{x|\omega},\label{eq:L-x-omg-expression}
\end{equation}
and $J_{\mu}(\sigma_{x}^{\text{SE}})$ can be therefore expressed
as
\begin{align}
J_{\mu}(\sigma_{x}^{\text{SE}}) & =\sum_{\omega}\left[K^{\text{cl}}\left(p(\omega|x)\right)\text{Tr}\left(\sigma_{x|\omega}E_{\mu}\right)+4\partial_{x}p(\omega|x)\text{ReTr}\left(\partial_{x}\sigma_{x|\omega}\sigma_{x|\omega}E_{\mu}\right)+4p(\omega|x)\text{Tr}\left(\partial_{x}\sigma_{x|\omega}\sigma_{x|\omega}\partial_{x}\sigma_{x|\omega}E_{\mu}\right)\right],
\end{align}
where 
\begin{equation}
\text{ReTr}\left(\partial_{x}\sigma_{x|\omega}\sigma_{x|\omega}E_{\mu}\right)=\text{Re}\braket{\psi_{x|\omega}\big|E_{\mu}\big|\partial_{x}\psi_{x|\omega}},
\end{equation}
\begin{equation}
\text{Tr}\left(\partial_{x}\sigma_{x|\omega}\sigma_{x|\omega}\partial_{x}\sigma_{x|\omega}E_{\mu}\right)=\braket{\partial_{x}\psi_{x|\omega}\big|E_{\mu}\big|\partial_{x}\psi_{x|\omega}}+\big|\braket{\partial_{x}\psi_{x|\omega}\big|\psi_{x|\omega}}\big|^{2}\braket{\psi_{x|\omega}\big|E_{\mu}\big|\psi_{x|\omega}}-2\text{Re}\braket{\psi_{x|\omega}\big|\partial_{x}\psi_{x|\omega}}\braket{\partial_{x}\psi_{x|\omega}\big|E_{\mu}\big|\psi_{x|\omega}}.
\end{equation}

On the other hand, the qCRB for single-parameter is always saturable~\citep{braunstein1994statistical,yang2019optimal},
i.e, there always exists a POVM measurement $\{E_{\mu}\}$ such that
Eq.~(\ref{eq:qCRB}) is saturated. As such, we conclude 
\begin{equation}
J_{\mu}^{\text{cl}}(\rho_{x};E_{\mu})\leq J_{\mu}(\rho_{x})\leq J_{\mu}(\sigma_{x}^{\text{SE}}),\label{eq:Refined-ConIneq-General}
\end{equation}
which we dub as refined convexity inequality. 

\section{Proof of Theorem~\ref{thm:lossless}}

We consider a pure parameter-dependent state $\ket{\Psi_{x}}$ and
a parameter-independent quantum channel $\{Q_{\omega}\}_{\omega\in\Omega}$.
In the perpendicular gauge, where 
\begin{equation}
\braket{\partial_{x}\Psi_{x}\big|\Psi_{x}}=0,
\end{equation}
so that $\ket{\partial_{x}^{\perp}\Psi_{x}}=\ket{\partial_{x}\Psi_{x}}$.
In this gauge, according to Table I in Ref.~\citep{yang2024theoryof},
we find the measurement inequality~(\ref{eq:Refined-Meas-Ineq})
in the main text is saturated if 
\begin{equation}
\text{Im}\braket{\partial_{x}\Psi_{x}\big|Q_{\omega}\big|\Psi_{x}}=0\label{eq:lossless-encoding-perp-gauge}
\end{equation}
When postselection is involved, saturating the postselection inequality~(\ref{eq:PS-ineq})
in the main text requires 
\begin{equation}
\braket{\partial_{x}\Psi_{x}\big|Q_{\omega}\big|\Psi_{x}}=0,\omega\in\checkmark,\label{eq:PS-saturate}
\end{equation}
and
\begin{equation}
\braket{\partial_{x}\Psi_{x}\big|Q_{\times}\big|\partial_{x}\Psi_{x}}=0,\label{eq:zero-discard}
\end{equation}
where 
\begin{equation}
Q_{\times}=\sum_{\omega\in\times}Q_{\omega}.
\end{equation}
Eq.~(\ref{eq:PS-saturate}) and (\ref{eq:zero-discard}) is the condition
for lossless compression. Furthermore, when Eq.~(\ref{eq:PS-saturate})
is saturated, the average QFI in the post-selected states is, according
to Ref.~\citep{yang2024theoryof} 
\begin{equation}
\sum_{\omega\in\checkmark}p(\omega|\checkmark)I(\sigma_{x|\omega})=4\sum_{\omega\in\checkmark}\braket{\partial_{x}^{\perp}\Psi_{x}|Q_{\omega}\big|\partial_{x}^{\perp}\Psi_{x}}=4\braket{\partial_{x}\Psi_{x}\big|Q_{\checkmark}\big|\partial_{x}\Psi_{x}}.\label{eq:ave-QFI}
\end{equation}
Now we take $Q_{\omega}=\mathbb{I}\otimes\Pi_{\omega}^{\text{E}}$
with $\Pi_{\omega}^{\text{E}}\equiv\ket{\pi_{\omega}^{\text{E}}}\bra{\pi_{\omega}^{\text{E}}}$
and define 
\begin{equation}
E_{\omega,\,x}\equiv M_{\omega,\,x}^{\dagger}M_{\omega,\,x},\,F_{\omega,\,x}\equiv\text{i}\partial_{x}M_{\omega,\,x}^{\dagger}M_{\omega,\,x},\,G_{\omega,\,x}\equiv\partial_{x}M_{\omega,\,x}^{\dagger}\partial_{x}M_{\omega,\,x}.
\end{equation}
It can be readily observed that

\begin{equation}
F_{\Omega,\,x}^{\dagger}=\sum_{\omega}F_{\omega,\,x}^{\dagger}=-\text{i}\sum_{\omega}M_{\omega,\,x}^{\dagger}\partial_{x}M_{\omega,\,x}=\sum_{\omega}F_{\omega,x}=F_{\Omega,\,x}.
\end{equation}
It is straightforward to calculate~\citep{escher2011general}
\begin{align}
\braket{\partial_{x}\Psi_{x}^{\text{SE}}\big|\partial_{x}\Psi_{x}^{\text{SE}}} & =\bra{\psi_{i}}\otimes\bra{\phi_{i}^{\text{E}}}\partial_{x}U_{x}^{\text{SE}\dagger}\partial_{x}U_{x}^{\text{SE}}\ket{\psi_{i}}\otimes\ket{\phi_{i}^{\text{E}}}\nonumber \\
 & =\sum_{\omega}\braket{\psi_{i}\big|\partial_{x}M_{\omega,\,x}^{\dagger}\partial_{x}M_{\omega,\,x}\big|\psi_{i}}=\langle G_{\Omega,\,x}\rangle,
\end{align}
\begin{align}
\braket{\partial_{x}\Psi_{x}^{\text{SE}}\big|\Psi_{x}^{\text{SE}}} & =\bra{\psi_{i}}\otimes\bra{\phi_{i}^{\text{E}}}\partial_{x}U_{x}^{\text{SE}\dagger}U_{x}^{\text{SE}}\ket{\psi_{i}}\otimes\ket{\phi_{i}^{\text{E}}}\nonumber \\
 & =\sum_{\omega}\braket{\psi_{i}\big|\partial_{x}M_{\omega,\,x}^{\dagger}M_{\omega,\,x}\big|\psi_{i}}=-\text{i}\langle F_{\Omega,\,x}\rangle.\label{eq:F-OMG-formula}
\end{align}
Thus, regardless of whether $M_{\omega}$ is lossless of not, 
\begin{equation}
I^{Q}=\langle G_{\Omega,\,x}\rangle-\left(\langle F_{\Omega,\,x}\rangle\right)^{2}.\label{eq:IQ-GF}
\end{equation}
In particular, in the perpendicular gauge, we find
\begin{equation}
\langle F_{\Omega,\,x}\rangle=\text{i}\langle\!\langle\partial_{x}U_{x}^{\text{SE}\dagger}U_{x}^{\text{SE}}\rangle\!\rangle=0,
\end{equation}
and 
\begin{equation}
I^{Q}=\langle G_{\Omega,\,x}\rangle.
\end{equation}
Furthermore, Eq.~(\ref{eq:lossless-encoding-perp-gauge}), Eq.~(\ref{eq:PS-saturate}),
and Eq.~(\ref{eq:zero-discard}) become 
\begin{equation}
\text{Re}\langle F_{\omega,\,x}\rangle=0,\,\forall\omega\in\Omega\label{eq:LL-enc-F}
\end{equation}
\begin{equation}
\langle F_{\omega,\,x}\rangle=0,\,\omega\in\checkmark,\label{eq:PS-FullQFI}
\end{equation}
\begin{equation}
\langle G_{\times,\,x}\rangle=0.\label{eq:Discard-ZeroQFI}
\end{equation}
Since Fur
\begin{align}
\text{Re}\langle F_{\omega,\,x}\rangle & =\frac{\text{i}}{2}\left(\langle\partial_{x}M_{\omega,\,x}^{\dagger}M_{\omega,\,x}\rangle-\langle M_{\omega,\,x}^{\dagger}\partial_{x}M_{\omega,\,x}\rangle\right),\\
\text{Im}\langle F_{\omega,\,x}\rangle & =\frac{1}{2}\left(\langle\partial_{x}M_{\omega,\,x}^{\dagger}M_{\omega,\,x}\rangle+\langle M_{\omega,\,x}^{\dagger}\partial_{x}M_{\omega,\,x}\rangle\right),
\end{align}
Eq.~(\ref{eq:LL-enc-F}) implies Eq.~~(\ref{eq:loss-encoding})
in the main text and Eq.~(\ref{eq:PS-FullQFI}) implies 
\begin{equation}
\langle\partial_{x}M_{\omega,\,x}^{\dagger}M_{\omega,\,x}\rangle=\langle M_{\omega,\,x}^{\dagger}\partial_{x}M_{\omega,\,x}\rangle=0,\omega\in\checkmark,
\end{equation}
i.e., Eq.~(\ref{eq:dMM-MdM}) in the main text. Eq.~(\ref{eq:Discard-ZeroQFI})
implies Eq.~(\ref{eq:dM-cross}) in the main text since $G_{\times,\,x}$
is semi-positive definite. 

\section{Theorem~\ref{thm:lossless} in the generic gauge}

For arbitrary given state $\ket{\Psi_{x}^{\text{SE}}}$ that does
not satisfy the perpendicular gauge condition, one can always make
the $U(1)$ gauge transformation 
\begin{align}
U_{x}^{\text{SE}} & \to\tilde{U}_{x}^{\text{SE}}e^{\text{i}\theta_{x}}\\
\ket{\Psi_{x}^{\text{SE}}} & \to\ket{\tilde{\Psi}_{x}^{\text{SE}}}=e^{\text{i}\theta_{x}}\ket{\Psi_{x}^{\text{SE}}},
\end{align}
such that $\ket{\tilde{\Psi}_{x}^{\text{SE}}}$ satisfies the perpendicular
gauge condition, i.e.,
\begin{equation}
\braket{\partial_{x}\tilde{\Psi}_{x}^{\text{SE}}\big|\tilde{\Psi}_{x}^{\text{SE}}}=\braket{\partial_{x}\Psi_{x}^{\text{SE}}\big|\Psi_{x}^{\text{SE}}}-\text{i}\partial_{x}\theta_{x}=0.
\end{equation}
which reduces to, according to Eq.~(\ref{eq:F-OMG-formula})
\begin{equation}
\partial_{x}\theta_{x}=-\langle F_{\Omega,\,x}\rangle.\label{eq:dtheta-dx}
\end{equation}
It is then straightforward to calculate that 
\begin{align}
\partial_{x}U_{x}^{\text{SE}\dagger}U_{x}^{\text{SE}} & \to\partial_{x}\tilde{U}_{x}^{\text{SE}\dagger}\tilde{U}_{x}^{\text{SE}}=\partial_{x}U_{x}^{\text{SE}\dagger}U_{x}^{\text{SE}}-\text{i}\partial_{x}\theta_{x}\\
\partial_{x}M_{\omega,\,x} & \to\partial_{x}\tilde{M}_{\omega,\,x}=e^{\text{i}\theta_{x}}\left(\partial_{x}M_{\omega,\,x}+\text{i}\partial_{x}\theta_{x}M_{\omega,\,x}\right),\\
F_{\omega,\,x} & \to\tilde{F}_{\omega,\,x}=F_{\omega,\,x}+\partial_{x}\theta_{x}E_{\omega,\,x},\\
G_{\omega,\,x} & \to\tilde{G}_{\omega,\,x}=G_{\omega,\,x}+\partial_{x}\theta_{x}(F_{\omega,\,x}+F_{\omega,\,x}^{\dagger})+(\partial_{x}\theta_{x})^{2}E_{\omega,\,x}.
\end{align}
In the generic gauge, Theorem~\ref{thm:lossless} becomes:
\begin{thm}
\textup{\label{thm:lossless-generic}A measurement encoding, which
does not necessarily satisfy the perpendicular gauge, is precision-preserving
if and only if 
\begin{equation}
\text{Re}\langle F_{\omega,\,x}\rangle=0,\,\omega\in\Omega.\label{eq:LL-enc-F-generic-gauge}
\end{equation}
Furthermore, when postselection is involved, it is required that}
\end{thm}
\begin{equation}
\langle F_{\omega,\,x}\rangle=\langle F_{\Omega,\,x}\rangle\langle E_{\omega,\,x}\rangle,\,\omega\in\checkmark,\label{eq:PS-FullQFI-Generic}
\end{equation}
\begin{equation}
\langle G_{\times,\,x}\rangle=\langle F_{\Omega,\,x}\rangle\langle F_{\times,\,x}\rangle,\label{eq:Discard-ZeroQFI-Generic}
\end{equation}

\begin{proof}
Upon applying Eqs.~(\ref{eq:LL-enc-F}-\ref{eq:Discard-ZeroQFI})
to $\tilde{U}_{x}^{\text{SE}}$, we obtain Eq.~(\ref{eq:LL-enc-F-generic-gauge})
and Eq.~(\ref{eq:PS-FullQFI-Generic}) immediately and 
\begin{equation}
\langle G_{\omega,\,x}\rangle+\partial_{x}\theta_{x}(\langle F_{\omega,\,x}\rangle+\langle F_{\omega,\,x}^{\dagger}\rangle)+(\partial_{x}\theta_{x})^{2}\langle E_{\omega,\,x}\rangle=0.\label{eq:Zero-QFI-generic-gauge-raw}
\end{equation}
To show the equivalence between Eq.~(\ref{eq:Zero-QFI-generic-gauge-raw})
and Eq.~(\ref{eq:Discard-ZeroQFI-Generic}), let us make a few observations:
(i) $\langle F_{\Omega,\,x}\rangle$, $\langle E_{\omega,\,x}\rangle$
and $\langle G_{\omega,\,x}\rangle$ are all real numbers. (ii) To
satisfy Eq.~(\ref{eq:PS-FullQFI-Generic}), it is necessary to have
\begin{align}
\text{Im}\langle F_{\omega,\,x}\rangle & =0,\:\omega\in\checkmark,\label{eq:ImF-check}\\
\text{Im}\langle F_{\times,\,x}\rangle & =\text{Im}\langle F_{\Omega,\,x}\rangle-\text{Im}\langle F_{\checkmark,\,x}\rangle=0.\label{eq:ImF-cross}
\end{align}
and thus we conclude Eq.~(\ref{eq:PS-FullQFI-Generic}) also implies
$\langle F_{\times,\,x}\rangle$ is real number. (iii) Eq.~(\ref{eq:PS-FullQFI-Generic})
also leads to 
\begin{equation}
\langle F_{\Omega,\,x}\rangle\langle E_{\times,\,x}\rangle=\langle F_{\Omega,\,x}\rangle-\langle F_{\Omega,\,x}\rangle\langle E_{\checkmark,\,x}\rangle=\langle F_{\times,\,x}\rangle.
\end{equation}
With these three observations, it is straightforward to obtain Eq.~(\ref{eq:Discard-ZeroQFI-Generic})
from Eq.~(\ref{eq:Zero-QFI-generic-gauge-raw}) and Eq.~(\ref{eq:dtheta-dx}).
\end{proof}
It is important to note that Eq.~(\ref{eq:LL-enc-F-generic-gauge})
is gauge invariant so is Eq.~(\ref{eq:loss-encoding}) in the main
text!

When only Eq.~(\ref{eq:PS-FullQFI-Generic}) is satisfied but Eq.~(\ref{eq:Discard-ZeroQFI-Generic})
may be violated, the QFI for the average post-selected state is, according
to Eq.~(\ref{eq:ave-QFI}), 
\begin{align}
\sum_{\omega\in\checkmark}p(\omega|x)I(\sigma_{x|\omega}) & =4\left(\langle G_{\checkmark,\,x}\rangle-\langle F_{\Omega,\,x}\rangle\langle F_{\checkmark,\,x}\rangle\right).\label{eq:ave-QFI-general}
\end{align}
Thus the loss of the QFI becomes 
\begin{equation}
\kappa=1-\frac{\langle G_{\checkmark,\,x}\rangle-\langle F_{\Omega,\,x}\rangle\langle F_{\checkmark,\,x}\rangle}{\langle G_{\Omega,\,x}\rangle-\langle F_{\Omega,\,x}\rangle^{2}}=\frac{\langle G_{\times,\,x}\rangle-\langle F_{\Omega,\,x}\rangle\langle F_{\times,\,x}\rangle}{\langle G_{\Omega,\,x}\rangle-\langle F_{\Omega,\,x}\rangle^{2}},\label{eq:kappa-general}
\end{equation}
which vanishes when Eq.~(\ref{eq:Discard-ZeroQFI-Generic}) is satisfied. 

\section{metrological information transduction using measurements: details
of the calculation}

Consider three stages (i) The ancillary system senses the parameter
of interest for some time $T$ with the Hamiltonian $H^{\text{E}}=xH_{0}^{\text{E}}$,
where its interaction with the target system is switched off. (ii)
Switch on the interaction between the ancillary system and the target
system (iii) a rank-$1$ projective measurements is performed on the
ancillary system, which implements. Without loss of generality, we
assume $\braket{\phi_{i}^{\text{E}}\big|H_{0}^{\text{E}}\big|\phi_{i}^{\text{E}}}=0$,
such that 
\begin{equation}
\text{Var}[H_{0}^{\text{E}}]_{\ket{\phi_{i}^{\text{E}}}}=\|H_{0}^{\text{E}}\ket{\phi_{i}^{\text{E}}}\|^{2}.
\end{equation}
At the end of stage (ii), it is readily to obtain 
\begin{equation}
U_{x}^{\text{SE}}=U_{\text{I}}e^{-\text{i}xH_{0}^{\text{E}}}.
\end{equation}
It is then straightforward to calculate
\begin{equation}
M_{\omega,x}=\braket{\pi_{\omega}^{\text{E}}|U_{x}^{\text{SE}}\big|\phi_{i}^{\text{E}}},\quad M_{\omega,x}=-\text{i}xT\braket{\pi_{\omega}^{\text{E}}|U_{x}^{\text{SE}}H_{0}^{\text{E}}\big|\phi_{i}^{\text{E}}}.
\end{equation}
Therefore, we find 
\begin{equation}
\braket{\tilde{\psi}_{x|\omega}|\partial_{x}\tilde{\psi}_{x|\omega}}=-\text{i}xT\bra{\psi_{i}}\otimes\braket{\phi_{i}^{\text{E}}|U_{x}^{\text{SE}\dagger}|\pi_{\omega}^{\text{E}}}\braket{\pi_{\omega}^{\text{E}}|U_{x}^{\text{SE}}H_{0}^{\text{E}}\big|\phi_{i}^{\text{E}}}\otimes\ket{\psi_{i}}.
\end{equation}
Upon using the weak signal approximation, $U_{x}^{\text{SE}}\approx U_{\text{I}}$
and noting that $\ket{\phi_{i}^{\text{E}\perp}}\equiv\frac{H_{0}^{\text{E}}\ket{\phi_{i}^{\text{E}}}}{\|H_{0}^{\text{E}}\ket{\phi_{i}^{\text{E}}}\|}$,
we find
\begin{equation}
\braket{\tilde{\psi}_{x|\omega}|\partial_{x}\tilde{\psi}_{x|\omega}}=-\text{i}x\sqrt{\text{Var}[H_{0}^{\text{E}}]_{\ket{\phi_{i}^{\text{E}}}}}T\bra{\psi_{i}}\otimes\braket{\phi_{i}^{\text{E}}|U_{\text{I}}^{\dagger}|\pi_{\omega}^{\text{E}}}\braket{\pi_{\omega}^{\text{E}}|U_{\text{I}}\big|\phi_{i}^{\text{E}\perp}}\otimes\ket{\psi_{i}}.
\end{equation}
Therefore, imposing Eq.~(\ref{eq:dMM-MdM}) in the main text leads
to
\begin{equation}
\bra{\psi_{i}}\otimes\braket{\phi_{i}^{\text{E}}|U_{\text{I}}^{\dagger}|\pi_{\omega}^{\text{E}}}\braket{\pi_{\omega}^{\text{E}}|U_{\text{I}}\big|\phi_{i}^{\text{E}\perp}}\ket{\psi_{i}}=0,\forall\omega\in\Omega,\label{eq:Opt-Enc-Transduc}
\end{equation}
 which is Eq.~(\ref{eq:Optimal-Encoding}) in the main text. 

Intuitively, if we can make
\begin{align}
U_{\text{I}}\ket{\psi_{i}}\otimes\ket{\phi_{i}^{\text{E}}} & \to\ket{\psi_{i}}\otimes\ket{\phi_{i}^{\text{E}}},\\
U_{\text{I}}\ket{\psi_{i}}\otimes\ket{\phi_{i}^{\perp\text{E}}} & \to\ket{\psi_{i}^{\perp}}\otimes\ket{\phi_{i}^{\perp\text{E}}},
\end{align}
where $\ket{\psi_{i}^{\perp}}$ is the state orthogonal to $\ket{\psi_{i}}$,
then Eq.~(\ref{eq:Opt-Enc-Transduc}) would be satisfied any rank-$1$
measurements. This motivates us to construct the control-$U$ gate~(\ref{eq:UI-expression})
in the main text. 

\section{The approximate completeness relation }

We show that the measurement operators for the discrete non-Hermitian
evolution satisfy the following property:
\begin{prop}
\textup{The measurement operators satisfy the approximate completeness
relation, i.e.,}
\end{prop}
\begin{equation}
\sum_{n=1}^{N}\sum_{j}M_{jn,\,x}^{\dagger}M_{jn,\,x}+M_{\checkmark,\,x}^{\dagger}M_{\checkmark,\,x}=\mathbb{I}+O(N\Delta t^{2}).\label{eq:weak-completenss}
\end{equation}

\begin{proof}
We prove by mathematical induction. It is straightforward to see that
for $N=2$, we have 
\begin{equation}
M_{j1,\,x}=M_{j}^{(1)},\,,M_{j2,\,x}=M_{j}^{(2)}M_{\checkmark,\,x}^{(1)},\,M_{\checkmark,\,x}=M_{\checkmark,\,x}^{(2)}M_{\checkmark,\,x}^{(1)}.
\end{equation}
Thus it is straightforward to check
\begin{align}
\sum_{n=1}^{2}\sum_{j}M_{jn,\,x}^{\dagger}M_{jn,\,x}+M_{\checkmark,\,x}^{\dagger}M_{\checkmark,\,x} & =\sum_{j}M_{j}^{(1)\dagger}M_{j}^{(1)}+\sum_{j}M_{\checkmark,\,x}^{(1)\dagger}M_{j}^{(2)\dagger}M_{j}^{(2)}M_{\checkmark,\,x}^{(1)}+M_{\checkmark,\,x}^{(1)\dagger}M_{\checkmark,\,x}^{(2)\dagger}M_{\checkmark,\,x}^{(2)}M_{\checkmark,\,x}^{(1)}\nonumber \\
 & =\sum_{j}M_{j}^{(1)\dagger}M_{j}^{(1)}+M_{\checkmark,\,x}^{(1)\dagger}\left(\sum_{j}M_{j}^{(2)\dagger}M_{j}^{(2)}+M_{\checkmark,\,x}^{(2)\dagger}M_{\checkmark,\,x}^{(2)}\right)M_{\checkmark,\,x}^{(1)}\nonumber \\
 & =\sum_{j}M_{j}^{(1)\dagger}M_{j}^{(1)}+M_{\checkmark,\,x}^{(1)\dagger}M_{\checkmark,\,x}^{(1)}\left(\mathbb{I}+O(\Delta t^{2})\right)\nonumber \\
 & =\mathbb{I}+O(2\Delta t^{2}).
\end{align}
Now assume Eq.~(\ref{eq:weak-completenss}) holds for $N$. Then
for $N+1$, we define 
\begin{align}
\tilde{M}_{\checkmark,\,x} & \equiv M_{\checkmark,\,x}^{(N+1)}M_{\checkmark,\,x},\;\tilde{M}_{jn,\,x}\equiv M_{jn,\,x},\;n\in[1,\,N],\;\tilde{M}_{jN+1,\,x}\equiv M_{j}^{(N+1)}M_{\checkmark,\,x},
\end{align}
where the variables with and without tildes correspond to the case
of $N+1$ and $N$ respectively. Then as before, it can be easily
checked that 
\begin{align}
\sum_{n=1}^{N+1}\sum_{j}\tilde{M}_{jn,\,x}^{\dagger}\tilde{M}_{jn,\,x}+\tilde{M}_{\checkmark,\,x}^{\dagger}\tilde{M}_{\checkmark,\,x} & =M_{\checkmark,\,x}^{\dagger}M_{\checkmark,\,x}^{(N+1)\dagger}M_{\checkmark,\,x}^{(N+1)}M_{\checkmark,\,x}+\sum_{j}M_{\checkmark,\,x}^{\dagger}M_{j}^{(N+1)\dagger}M_{j}^{(N+1)}M_{\checkmark,\,x}+\sum_{j}\sum_{n=1}^{N}M_{jn,\,x}^{\dagger}M_{jn,\,x}\nonumber \\
 & =M_{\checkmark,\,x}^{\dagger}\left(\mathbb{I}+O(\Delta t^{2})\right)M_{\checkmark,\,x}+\sum_{j}\sum_{n=1}^{N}M_{jn,\,x}^{\dagger}M_{jn,\,x}\nonumber \\
 & =\mathbb{I}+O((N+1)\Delta t^{2}),
\end{align}
which completes the proof.
\end{proof}
In the continuum limit where $\Delta t\to0$, $N\to\infty$ and $N\Delta t\to T$
, the approximate completeness relation becomes the integral completeness
relation:
\begin{align}
\sum_{n=1}^{N}\sum_{j}M_{jn,\,x}^{\dagger}M_{jn,\,x}+M_{\checkmark,\,x}^{\dagger}M_{\checkmark,\,x} & =\sum_{j}\sum_{n=1}^{N}\gamma_{j}(t_{n})M_{\checkmark,\,x}^{(1)\dagger}M_{\checkmark,\,x}^{(n-1)\dagger}L_{j}^{\dagger}L_{j}M_{\checkmark,\,x}^{(n-1)}\cdots M_{\checkmark,\,x}^{(1)}\Delta t+M_{\checkmark,\,x}^{\dagger}M_{\checkmark,\,x}\nonumber \\
 & =\sum_{j}\int_{0}^{T}\gamma_{j}(t)M_{\checkmark,\,x}^{\dagger}(t)L_{j}^{\dagger}L_{j}M_{\checkmark}(t)dt+M_{\checkmark,\,x}^{\dagger}(T)M_{\checkmark,\,x}(T)=\mathbb{I}.
\end{align}

\section{Proof of Theorem~\ref{thm:Lossles-NH-sensing}}

We consider a sequence of unitary encoding interspersed by post-selections
\begin{equation}
U_{x}^{\text{SE}}=U_{x}^{\text{SE}(N)}\cdots U{}_{x}^{\text{SE}(2)}U_{x}^{\text{SE}(1)},
\end{equation}
as shown in Fig.~\ref{fig:NH-Sensing}(b) in the main text. In the
discrete case where $T=N\Delta t$, we note that the operators corresponding
to the retained state
\begin{equation}
M_{\checkmark,\,x}=M_{\checkmark,\,x}^{(N)}\cdots M{}_{\checkmark,\,x}^{(2)}M_{\checkmark,\,x}^{(1)},
\end{equation}
and operators correspond to the discarded states are
\begin{equation}
M_{jn,\,x}=M_{j}^{(n)}M_{\checkmark,\,x}^{(n-1)}\cdots M_{\checkmark,\,x}^{(2)}M_{\checkmark,\,x}^{(1)},
\end{equation}
where $n=1,\,2,\cdots N$. 

Now we are in a position to prove Theorem~\ref{thm:Lossles-NH-sensing}.
Let us note that under the gauge transformation,
\begin{align}
U_{x}^{\text{SE}} & \to\bar{U}_{x}^{\text{SE}}=e^{\text{i}\theta_{x}}U_{x}^{\text{SE}},\\
M_{\omega,\,x} & \to\bar{M}_{\omega,\,x}=e^{\text{i}\theta_{x}}M_{\omega,\,x},
\end{align}
one can readily check
\begin{align}
\partial_{x}U_{x}^{\text{SE}\dagger}U_{x}^{\text{SE}} & \to\partial_{x}\bar{U}_{x}^{\text{SE}\dagger}\bar{U}_{x}^{\text{SE}}=\partial_{x}U_{x}^{\text{SE}\dagger}U_{x}^{\text{SE}}-\text{i}\partial_{x}\theta_{x},\\
\partial_{x}M_{\omega,\,x} & \to\partial_{x}\bar{M}_{\omega,\,x}=e^{\text{i}\theta_{x}}\left(\partial_{x}M_{\omega,\,x}+\text{i}\partial_{x}\theta_{x}M_{\omega,\,x}\right),\\
F_{\omega,\,x} & \to\bar{F}_{\omega,\,x}=F_{\omega,\,x}+\partial_{x}\theta_{x}E_{\omega,\,x}.
\end{align}
Clearly, $E_{\omega,\,x}$ is gauge-invariant. Then one can easily
check for $\omega\in\checkmark$
\begin{align}
\langle\bar{F}_{\omega,\,x}\rangle & =\langle F_{\omega,\,x}\rangle+\partial_{x}\theta_{x}\langle E_{\omega,\,x}\rangle,
\end{align}

For non-Hermitian sensing, we have only one retained state. Without
causing any ambiguity, we can simply use $\checkmark$ instead of
$\omega\in\checkmark$. Based on the above observation, we can gauge
\begin{align}
M_{\checkmark,\,x} & \to\bar{M}_{\checkmark,\,x}=M_{\checkmark,\,x}e^{\text{i}\theta_{\checkmark,\,x}},\\
M_{jn,\,x} & \to\bar{M}_{jn,\,x}=M_{jn,\,x}e^{\text{i}\theta_{\checkmark,\,x}},
\end{align}
such that 
\begin{equation}
\text{Re}\langle\bar{F}_{\checkmark,\,x}\rangle=0,
\end{equation}
where $\theta_{\checkmark,\,x}$ satisfies
\begin{equation}
\partial_{x}\theta_{\checkmark,\,x}\langle E_{\checkmark,\,x}\rangle=-\text{Re}\langle F_{\checkmark,\,x}\rangle.
\end{equation}
Since lossless measurement encoding must satisfy Theorem~\ref{thm:lossless-generic},
which implies Eq.~(\ref{eq:ImF-check}) must satisfied, i.e..
\begin{equation}
\text{Im}\langle\bar{F}_{\checkmark,\,x}\rangle=\frac{1}{2}\partial_{x}\langle E_{\checkmark,\,x}\rangle=0,
\end{equation}
and therefore 
\begin{equation}
\langle\bar{F}_{\checkmark,\,x}\rangle=0.
\end{equation}
Furthermore, since $\langle E_{\checkmark,\,x}\rangle$ is strictly
positively, then Eq.~(\ref{eq:PS-FullQFI-Generic}) forces 
\begin{equation}
\langle\bar{F}_{\Omega,\,x}\rangle=0,
\end{equation}
which implies $\ket{\bar{\Psi}_{x}^{\text{SE}}}=\ket{\Psi_{x}^{\text{SE}}}e^{\text{i}\theta_{\checkmark,\,x}}$
satisfies the perpendicular gauge condition. This condition in turn
leads to, according to Eq.~(\ref{eq:Discard-ZeroQFI-Generic}), 
\begin{equation}
\langle\bar{G}_{\times,\,x}\rangle=0,
\end{equation}
which is equivalent to
\begin{equation}
\partial_{x}\bar{M}_{jn,\,x}\ket{\psi_{i}}=0.
\end{equation}
As a result, we know 
\begin{equation}
\langle\bar{F}_{jn,\,x}\rangle=0,
\end{equation}
and thus $\langle\bar{F}_{\Omega,\,x}\rangle=0$. Then it straightforward
to calculate 
\begin{equation}
\partial_{x}\bar{M}_{jn,\,x}=\left(\partial_{x}M_{jn,\,x}+\text{i}\partial_{x}\theta_{\checkmark,\,x}M_{jn,\,x}\right)e^{\text{i}\theta_{\checkmark,\,x}},
\end{equation}
\begin{equation}
\partial_{x}\bar{M}_{jn,\,x}\ket{\psi_{i}}=\sqrt{\gamma_{j}(t_{n})\Delta t}L_{j}\left(\ket{\partial_{x}\tilde{\psi}_{x|\checkmark}(t_{n-1})}+\text{i}\partial_{x}\theta_{\checkmark,\,x}\ket{\tilde{\psi}_{x|\checkmark}(t_{n-1})}\right)=\sqrt{\gamma_{j}(t_{n})\Delta t}L_{j}\ket{\tilde{\xi}_{\checkmark,\,x}(t_{n-1})},
\end{equation}
where 
\begin{equation}
\ket{\tilde{\psi}_{x|\checkmark}(t_{n-1})}\equiv M_{\checkmark,\,x}^{(n-1)}\cdots M_{\checkmark,\,x}^{(2)}M_{\checkmark,\,x}^{(1)}\ket{\psi_{i}}.
\end{equation}
Thus 
\begin{equation}
\langle\bar{G}_{\times,\,x}\rangle=\sum_{j}\sum_{n=1}^{N}\langle\partial_{x}\bar{M}_{jn,\,x}^{\dagger}\partial_{x}\bar{M}_{jn,\,x}\rangle=\sum_{j}\sum_{n=1}^{N}\gamma_{j}(t_{n})\braket{\tilde{\xi}_{\checkmark,\,x}(t_{n-1})\big|L_{j}^{\dagger}L_{j}\big|\tilde{\xi}_{\checkmark,\,x}(t_{n-1})}\Delta t.
\end{equation}
In the continuum limit, 
\begin{align}
\langle\bar{G}_{\times,\,x}\rangle & =\sum_{j}\int_{0}^{T}\gamma_{j}(t)\braket{\tilde{\xi}_{\checkmark,\,x}(t)\big|L_{j}^{\dagger}L_{j}\big|\tilde{\xi}_{\checkmark,\,x}(t)}dt\nonumber \\
 & =\sum_{j}\int_{0}^{T}\gamma_{j}(t)\left[\braket{\partial_{x}\tilde{\psi}_{x|\checkmark}(t)|L_{j}^{\dagger}L_{j}|\partial_{x}\tilde{\psi}_{x|\checkmark}(t)}+(\partial_{x}\theta_{\checkmark,\,x})^{2}\braket{\tilde{\psi}_{x|\checkmark}(t)|L_{j}^{\dagger}L_{j}|\tilde{\psi}_{x|\checkmark}(t)}+2\partial_{x}\theta_{\checkmark,\,x}\text{Im}\braket{\tilde{\psi}_{x|\checkmark}(t)|L_{j}^{\dagger}L_{j}|\partial_{x}\tilde{\psi}_{x|\checkmark}(t)}\right]dt,
\end{align}
Similarly, we find 
\begin{equation}
\langle\bar{F}_{\times,\,x}\rangle=\text{i}\sum_{j}\sum_{n=1}^{N}\langle\partial_{x}\bar{M}_{jn,\,x}^{\dagger}\bar{M}_{jn,\,x}\rangle=\text{i}\sum_{j}\int_{0}^{T}\gamma_{j}(t)\braket{\tilde{\xi}_{\checkmark,\,x}(t)\big|L_{j}^{\dagger}L_{j}\big|\tilde{\psi}_{\checkmark,\,x}(t)}dt.
\end{equation}

\section{Integral representations of $\langle G_{\Omega}\rangle$ and $\langle F_{\Omega}\rangle$
for the quantum collision model}

Similar to the integral representations above, it can be readily found,
\begin{align}
\langle G_{\Omega,x}\rangle & =\langle\partial_{x}M_{\checkmark,\,x}^{\dagger}\partial_{x}M_{\checkmark,\,x}\rangle+\sum_{j}\sum_{n=1}^{N}\langle\partial_{x}M_{jn,\,x}^{\dagger}\partial_{x}M_{jn,\,x}\rangle\nonumber \\
 & =\langle G_{\checkmark,\,x}\rangle+\sum_{j}\int_{0}^{T}\gamma_{j}(t)\langle\partial_{x}M_{\checkmark,\,x}^{\dagger}(t)L_{j}^{\dagger}L_{j}\partial_{x}M_{\checkmark,\,x}(t)\rangle dt,
\end{align}
and 
\begin{align}
\langle F_{\Omega,x}\rangle & =\text{i}\langle\partial_{x}M_{\checkmark,\,x}^{\dagger}M_{\checkmark,\,x}\rangle+\text{i}\sum_{j}\sum_{n=1}^{N}\langle\partial_{x}M_{jn,\,x}^{\dagger}M_{jn,\,x}\rangle\nonumber \\
 & =\langle F_{\checkmark,\,x}\rangle+\text{i}\sum_{j}\int_{0}^{T}\gamma_{j}(t)\langle\partial_{x}M_{\checkmark,\,x}^{\dagger}(t)L_{j}^{\dagger}L_{j}M_{\checkmark,\,x}(t)\rangle dt.
\end{align}
According to Eq.~(\ref{eq:IQ-GF}), one can compute the $I^{Q}$
for the quantum collision model and hence the loss of the average
QFI $\kappa$.
\end{document}